\documentclass[a4paper,showpacs,nofootinbib]{revtex4-1}
\usepackage[utf8]{inputenc}
\usepackage{graphicx,bm,color,amssymb,amsmath}
\usepackage{slashed,verbatim}
\usepackage{subfigure}
\usepackage[colorlinks=true,linktocpage=true,linkcolor=blue,citecolor=blue]{hyperref}
\usepackage[font=small,labelfont=bf]{caption}
\usepackage[toc,page]{appendix}

\begin{document}
\title{An NJL model analysis of a magnetised nonextensive QCD medium}
\author{Chowdhury Aminul Islam}
\affiliation{Institut f\"{u}r Theoretische Physik, Johann Wolfgang Goethe–Universit\"{a}t, Max-von-Laue-Str. 1, D–60438 Frankfurt am Main, Germany}
\affiliation{School of Nuclear Science and Technology, University of Chinese Academy of Sciences, Beijing 100049, China}
\email{chowdhury.aminulislam@gmail.com}
\email{chowdhury@physik.uni-frankfurt.de}

\begin{abstract}
{We investigate the effect of a background magnetic field when applied to a nonextensive QCD medium in a $2+1$ flavour Nambu\textemdash Jona-Lasinio model. The effect of a constant as well as an $eB$-dependent coupling is considered. For the constant coupling, the well-known magnetic catalysis effect is observed with reduced strength in both the condensates and the transition temperatures compared to the standard extensive medium. In the case of a field-dependent coupling, we observe a competition between the nonextensive parameter, $q$ and $eB$. With sufficiently high $q$-values, the phenomenon of inverse magnetic catalysis \textemdash\, a well-known trait in effective models featuring $eB$-dependent coupling, appears to be eliminated within the range of magnetic fields examined in our present study.}
\end{abstract}

\maketitle

\section{Introduction} 
\label{sec:intr}
The creation of a strong magnetic field in heavy-ion collisions (HIC) is now an established fact. Its strength in non-central collisions can be huge as compared to any known terrestrial or extraterrestrial magnetic field-generating phenomenon. Depending on the collisional energy the strength of the magnetic field can vary between $\sim 1-15\, m_\pi^2$ from  Relativistic Heavy Ion Collider (RHIC) to Large Hadron Collider (LHC)~\cite{Skokov:2009qp}.

From the later half of the first decade in this century, the effect of the magnetic field on the properties of the QCD medium have been explored extensively using both theoretical and experimental techniques~\cite{Kharzeev:2013jha,Miransky:2015ava}. Numerous novel phenomena have been proposed and studied in that context, some of the most studied ones are magnetic catalysis (MC)~\cite{Gusynin:1994re}, inverse magnetic catalysis (IMC)~\cite{Bali:2012zg,Bali:2011qj}, chiral magnetic effect~\cite{Fukushima:2008xe}, chiral separation effect~\cite{Son:2004tq,Metlitski:2005pr} etc. Those interested can delve into an engaging review article that examines a spectrum of potential novel non-dissipative phenomena in the presence of magnetic fields and beyond~\cite{Kharzeev:2015znc}.

On the other hand, in the pursuit of understanding the data from HIC experiments both in RHIC and LHC one statistical distribution, recently, has gained a lot of attention~\cite{PHENIX:2011rvu,ALICE:2011gmo,CMS:2011jlm,Patra:2020gzw,Pradhan:2023bld}. This is widely known as the Tsallis distribution after Constantino Tsallis who in his famous paper~\cite{Tsallis:1987eu} investigated the possibility of a nonextensive form of entropy involving a parameter $q$. The extensive nature of the entropy is obtained back in the limit $q\rightarrow1$. As compared to the exponentials in traditional Boltzmann-Gibbs (BG) extensive statistics, the nonextensive statistics lead to power laws. Its application is found in almost every branch of physics~\cite{Tsallis:2009zex}.

Its applicability to describe the fireball created in HICs stems from the following facts. The fireball is finite, rapidly evolving in nature and also experiences strong intrinsic fluctuations and long-range correlations. These characteristics render the system non-uniform and global equilibrium cannot be established. As a consequence, some quantities develop power law tailed behaviour and the nonextensive statistics becomes applicable.

There are multiple versions of the Tsallis distribution available in the literature~\cite{Tsallis:1987eu,Tsallis:1998ws,Biro:2008hz,Wilk:2008ue}. Theoretical issues concerning the thermodynamic consistency while implementing Tsallis distribution in the case of relativistic high energy distribution~\cite{Pereira:2009ja,Conroy:2010wt} have been clarified in Ref.~\cite{Cleymans:2011in,Cleymans:2012ya}. A recent review on the topic concerning nuclear and particle physics can be found in Ref.~\cite{Kapusta:2021zfo}.

There have also been multiple developments in applying nonextensive statistics in QCD-based effective model describing both the hadronic and quark degrees of freedoms~\cite{Pereira:2007hp,Rozynek:2009zh,Rozynek:2015zca,Zhao:2020xob,Pradhan:2021vtp,Zhao:2023xpj}. These studies, among other things, discussed scalar and vector meson fields, properties of $\pi$ and $\sigma$ mesons, chiral symmetry restoration, effect of background gauge field (Polyakov loop field), characteristics of magnetisation, thermodynamic properties of QCD medium etc. 

Existing literatures on utilising the effective models for a nonextensive QCD medium do not incorporate the background magnetic field~\cite{Rozynek:2009zh,Rozynek:2015zca,Zhao:2020xob,Zhao:2023xpj}. On the other hand, there have been ample efforts to understand the effects of MC, IMC and magnetic properties like magnetisation, magnetic susceptibility, etc., of the QCD medium in different effective models~\cite{Farias:2014eca,Ferreira:2014kpa,Ayala:2014gwa,Farias:2016gmy,Tawfik:2016lih,Tawfik:2016gye,Tawfik:2017cdx,Tawfik:2021eeb}, albeit using extensive statistics. Thus, there is currently a gap in the existing literature regarding effective models for a nonextensive QCD medium that is also magnetized.

Keeping this in mind, in the present work, we shall present a nonextensive version of the Nambu\textemdash Jona-Lasinio (NJL) model describing a magnetised QCD medium. Specifically, chiral symmetry restoration while intertwined with nonextensiveness and magnetic field will occupy the central stage of our investigation. Such a study, besides addressing the existing theoretical gap, could have relevance for understanding magnetized QCD media with strong intrinsic fluctuations, long-range interactions, etc., potentially encountered in heavy-ion collision experiments. We start our investigation for the nonextensive statistics with a zero magnetic field $(eB)$ for $2+1$ flavours to put things into perspective. Then we move to the case of nonzero strength of $eB$. To gauge the effect of the nonextensive statistics we execute the analysis for BG statistics in parallel. We perform the analysis in presence of $eB$ considering different possible scenarios.

First, we check results for the standard NJL model where the coupling constant is independent of $eB$. Our result shows that at a specific magnetic field strength, the impact of increasing q resembles that of the zero magnetic field case. It leads to a decreasing trend in both the condensates and the transition temperatures. We find out that the model produces the MC effect in Tsallis statistics as well. However, it results in reduced strength in both the condensates and the transition temperatures compared to the BG statistics.

Then, we take our analysis to the case of a magnetic field dependent coupling constant. We choose to work with a coupling constant whose dependence on $eB$ has been determined by fitting to the available lattice QCD (LQCD) data, which produces the IMC effect~\cite{Ferreira:2014kpa}. We discover that there is a competition between the two parameters $q$ and $eB$. This is one of the most important findings of the present analysis. Depending on their strengths the observed IMC effect in BG statistics can be eliminated.

Our findings also suggest that with $eB$-dependent $G_S$ the trend of the transition temperature is non-monotonic. For smaller values of $q$, such as $q=1.05$, it remains constant initially and then starts decreasing with increasing magnetic field. With further increment in $q$, it exhibits a non-monotonic trend as a function of $eB$: initially it increases and then starts falling. To render the present analysis complete, we also explore the basic form of the coupling constant inspired by Ref.~\cite{Miransky:2002rp}, in which the strong coupling $\alpha_s$ decreases with increasing $eB$ and put it in the appendix. 

The remainder of the manuscript is organised as follows: First, we provide a brief review on the formalism for $2+1$ flavour NJL model in section~\ref{sec:for}. We start with the standard NJL model in BG statistics in subsection~\ref{ssec:stnd_stat} and then extend it to the nonextensive statistics in subsection~\ref{ssec:tsa_stat}. In section~\ref{sec:res}, we thoroughly discuss our findings and compare them with the known results for detailed analysis and comparison. Finally, in section~\ref{sec:con}, we summarise the conclusion.

\section{Formalism}
\label{sec:for}

\subsection{Extensive statistics}
\label{ssec:stnd_stat}
In this section, we briefly revisit the NJL model in the standard extensive scenario. The $2+1$ flavour NJL Lagrangian in presence of a magnetic field can be written as~\cite{Menezes:2008qt,Boomsma:2009yk,Chatterjee:2011ry,Avancini:2011zz,Farias:2014eca,Ferrer:2014qka,Yu:2014xoa,Mao:2016fha,Farias:2016gmy},
\begin{align}
\mathcal{L}_{\rm{NJL}}^B = \bar{\psi}(i\slashed D-m_0)\psi
+ \mathcal{L}_1+\mathcal{L}_2-\frac{1}{4}F^{\mu\nu}F_{\mu\nu},
\label{eq:lag_njl}
\end{align}
where $m_0\,=\, \mathrm{diag}(m_u,m_d,m_s)$ with $m_u\,=\,m_d$; $\slashed D=\gamma_\mu D^\mu$ with $D^\mu=\partial^\mu-iqA^\mu$ and $q$ being the electric charge ($q_u=2/3e$, $q_d=-1/3e$ and $q_s=-1/3e$); $e$ is the charge of a proton, $F^{\mu\nu}=\partial^\mu A^\nu-\partial^\nu A^\mu$ is the field strength tensor and $A^\mu$ being the electromagnetic gauge field. The Lagrangians $\mathcal{L}_1$ and $\mathcal{L}_2$ are~\cite{Klevansky:1992qe,Hatsuda:1994pi,Buballa:2003qv}
\begin{align}
 \mathcal{L}_1=&\frac{G_S}{2}\sum_{a=0}^{8}[(\bar{\psi}\lambda_a\psi)^2+(\bar{\psi}i\gamma_5\lambda_a\psi)^2]\,\,\mathrm{and}\\
 \mathcal{L}_2=&-G_D\{\mathrm{det}[\bar{\psi}(1+\gamma_5)\psi]+\mathrm{det}[\bar{\psi}(1-\gamma_5)\psi]\}.
\end{align}
The Lagrangian $\mathcal{L}_1$ is usually called the symmetric Lagrangian as it respects the full global symmetry of QCD, $SU(3)_V\times SU(3)_A\times U(1)_V\times U(1)_A$. On the other hand, $\mathcal{L}_2$ breaks the $U(1)_A$ symmetry and is responsible for the axial anomaly found in QCD. It is known as the 't Hooft determinant term or simply the determinant term. 

Before obtaining the thermodynamic potential, one must note a few important things. Because of the presence of the magnetic field there are some important modifications in the calculation. First, the dispersion relation of the quarks will be modified as,
\begin{align}
E_f(B)=[M_f^2+p_z^2+(2l+1-s)\lvert q_f\rvert B]^{1/2},
\label{eq:disp_mag}
\end{align}
where $M_f$ is the effective quark mass for a quark of flavour $f$. The magnetic field is in the $z$ direction $(\vec B=B\hat z)$, which results in discretising the momentum in the transverse plane, here in the $x-y$ plane. The transverse component of the momentum now depends on both the Landau level (LL), $l$ and the spin, $s$. It is important to note how in the magnetised version the energy of the individual flavour is dependent not only on the respective effective masses but also on the respective charges. Another important change, which is somewhat expected from Eq.~\ref{eq:disp_mag}, is in the integral over the three momenta,
\begin{align}
\int\frac{d^3p}{(2\pi)^3}\rightarrow\frac{|q_f|B}{2\pi}\sum_{l=0}^{\infty}\int_{-\infty}^{\infty} \frac{dp_z}{2\pi}.
\end{align}
The integral is now performed only over the continuous component, $p_z$. Also, note the change in the lower limit of the integral as compared to the zero magnetic field case, where it is $0$.

With all these knowledge in hand, we can get the thermodynamic potential potential for the NJL model in presence of the magnetic field as,
\begin{align}
 \Omega_{\mathrm{NJL}}^B(T,\mu)= G_S\sum_{f=u,d,s}\sigma_f^2-4G_D\sigma_u\sigma_d\sigma_s+\sum_{f=u,d,s}\left(\Omega_{\mathrm{mev}}^f+\Omega_{\mathrm{med}}^f\right)+\frac{B^2}{2},
\end{align}
where, $\sigma_f=\langle\bar\psi_f\psi_f\rangle$ with $f=u,\,d\,\mathrm{and}\,s$\footnote{Please note that the second term in the potential has a opposite sign in Ref.~\cite{Ferreira:2014kpa}. That's a typo. Our sign matches with Ref.~\cite{Rehberg:1995kh,Buballa:2003qv}}. The first two terms are the vacuum contributions from the condensates of different flavours. The second term arises from the 't Hooft determinant term and is the explicit manifestation of preservation of the axial anomaly. $\Omega_{\mathrm{mev}}^f$ is a term which we call magnetic field entangled vacuum term. It can be separated into a purely magnetic field dependent term and the standard vacuum term, which we show in a moment. $\Omega_{\mathrm{med}}^f$ is the term in presence of both the magnetic field and the temperature and can be termed as the medium term.

These two terms are given as,
\begin{align} 
\Omega_{\mathrm{mev}}^f=-\frac{N_c}{2\pi}\sum_{f,l,s}|q_f|B\int_{-\infty}^{\infty}\frac{dp_z}{(2\pi)}E_f(B)\;\;\; \mathrm{and} 
\end{align}
\begin{align}
\Omega_{\mathrm{med}}^f=-\frac{N_c}{2\pi}T\sum_{f,l,s}|q_f|B\int_{-\infty}^{\infty}\frac{dp_z}{(2\pi)} \left[{\rm{ln}}\left(1+e^{-(E_f(B)-{\mu})/T}\right)+{\rm{ln}}\left(1+e^{-(E_f(B)+{\mu})/T}\right) \right],
\label{eq:pot_njl_med}
\end{align}
respectively. $f$, $l$ and $s$ stand for the summation over flavour, Landau levels and spin states of the quarks, respectively. Following the technique in Ref.~\cite{Menezes:2008qt}, $\Omega_{\mathrm{mev}}^f$ can be further decomposed into the standard vacuum term $(\Omega_{\mathrm{vac}}^f)$ and a purely magnetic field dependent term $(\Omega_{\mathrm{mag}}^f)$, $\Omega_{\mathrm{mev}}^f\,=\,\Omega_{\mathrm{vac}}^f+\Omega_{\mathrm{mag}}^f$, as
\begin{align}
 \Omega_{\mathrm{vac}}^f = -2N_c\int_{\Lambda}\frac{d^3p}{(2\pi)^3}E_p^f\;\;\; \mathrm{and}
\end{align}
\begin{align}
 \Omega_{\mathrm{mag}}^f =-\frac{N_c}{2\pi^2}\sum_{f}(|q_f|B)^2\big(\zeta^{\prime}(-1,x_f)+\frac{x^2_f}{4}-\frac{1}{2}(x^2_f-x_f){\rm ln}x_f\big),
\end{align}
where $E_p^f=\sqrt{M_f^2+p^2}$ is the particle energy in vacuum\footnote{One should note that just by putting $B=0$ in Eq.~\ref{eq:disp_mag} the vacuum dispersion relation cannot be recovered.}, $x_f=M_f^2/(2|q_f|B)$ and $\zeta^{\prime}(-1,x_f)=d\zeta(z,x_f)/dz|_{z=-1}$ with $\zeta(z,x_f)$ is the Riemann-Hurwitz zeta function. The detail of obtaining the expression for $\Omega_{\mathrm{mag}}$ can be found in Ref.~\cite{Menezes:2008qt}. 

Thus, the final expression for the thermodynamic potential in NJL model becomes,
\begin{align}
 \Omega_{\mathrm{NJL}}^B(T,\mu)= G_S\sum_{f=u,d,s}\sigma_f^2-4G_D\sigma_u\sigma_d\sigma_s+\sum_{f=u,d,s}\left(\Omega_{\mathrm{vac}}^f+\Omega_{\mathrm{mag}}^f+\Omega_{\mathrm{med}}^f\right)+\frac{B^2}{2}.
 \label{eq:pot_njl_mag_final}
\end{align}

We take the model parameters as: Number of light flavours $= 2$, number of colours $=3$, $m_u=m_d=0.0055$ GeV, $m_s=0.1407$ GeV, three momentum cut-off $\Lambda=0.6023$ GeV, the scalar coupling constant\footnote{Note that the value of $G_S$ quoted here or in Ref.~\cite{Ferreira:2014kpa} is double the value given in Ref.~\cite{Rehberg:1995kh,Buballa:2003qv}. This is because of the the definition of the coupling constant $G_S$. It appears in the Lagrangian here (or in Ref.~\cite{Ferreira:2014kpa}) with a factor of $1/2$ in contrast to just $G_S$ in Ref.~\cite{Rehberg:1995kh,Buballa:2003qv}.}, $G_S=3.67/\Lambda^2$, the coupling constant for the 't Hooft determinant term, $G_D=12.36/\Lambda^5$, pion mass $m_\pi=0.135$ GeV and the pion decay constant, $f_\pi=0.0924$ GeV ($0.0879$ GeV, in the chiral limit), kaon mass $m_k=0.4977$ GeV, eta mesons, $m_\eta^{'}=0.9578$ GeV and $m_\eta=0.5148$ GeV. All these parameters are taken from the Ref.~\cite{Rehberg:1995kh}.

The magnetic field dependent coupling constant is fitted as~\cite{Ferreira:2014kpa}, 
\begin{align}
 G_S(\xi)=G_S^0\frac{1+a\xi^2+b\xi^3}{1+c\xi^2+d\xi^4},
 \label{eq:coupling_eB}
\end{align}
where, $a = 0.0108805$, $b = -1.0133\times10^{-4}$, $c = 0.02228$, $d = 1.84558\times10^{-4}$; $\Lambda_{\mathrm QCD}=0.3$ GeV, $\xi=eB/\Lambda_{\mathrm QCD}^2$ and $G_S^0=G_S(eB=0)=G_S$.

Because of the 't Hooft determinant term there is a cross-flavour coupling which is called ``flavour mixing''~\cite{Buballa:2003qv}. As a result the effective masses are given by
\begin{align}
M_u=&m_u-2G_S\sigma_u+2G_D\sigma_d\sigma_s\,,\nonumber\\
M_d=&m_d-2G_S\sigma_d+2G_D\sigma_u\sigma_s\;\;\;\;\; {\rm and}\nonumber\\
M_s=&m_s-2G_S\sigma_s+2G_D\sigma_u\sigma_d,
\label{eq:eff_mass}
\end{align}
for up, down and strange quarks, respectively.

\subsection{Nonextensive statistics}
\label{ssec:tsa_stat}
We know that the entropy is considered to be an extensive property, i.e., its value depends on the system size. Constantino Tsallis proposed a nonextensive statistics~\cite{Tsallis:1987eu}, which provides a generalisation of the traditional BG entropy. This generalisation is made through a parameter $q$, which, in the limit $q\rightarrow1$, gives back the BG entropy.

As compared to the exponentials in traditional extensive statistics, the nonextensive statistics lead to power laws. This is the most fundamentally different manifestation of the entropy being a nonextensive quantity.

Mathematically speaking, in nonextensive statistics or the so-called Tsallis statistics, the usual exponential are replaced by an equivalent q-exponential,
\begin{align}
 {\rm exp}\left(-\frac{E}{T}\right)\rightarrow {\rm exp}_{q}\left(-\frac{E}{T}\right),
\end{align}
where, 
\begin{align}
 {\rm exp}_q(x)\equiv\begin{cases}
                      \big(1+(q-1)x\big)^{1/(q-1)},\;\; \text{if $x>0$}\\
                      \big(1+(1-q)x\big)^{1/(1-q)},\;\; \text{if $x\leq0$}.
                     \end{cases}
 \label{eq:non_ext_exp}
\end{align}
Using these expressions one can get the generalised versions of the standard statistical distributions. One can derive the Tsallis distribution starting from the Boltzmann equation~\cite{Biro:2005uv}. It is straightforward to check that the above equation reduces to the standard exponential in the limit $q\rightarrow1$ giving back the standard distributions. Any values of $q$ other than unity gives a measure of the ``nonextensiveness'' of the system. In the same fashion the inverse function, i.e., the equivalent to standard logarithm can as well be defined,
\begin{align}
 {\rm ln}_q(x)\equiv\begin{cases}
                      \frac{x^{q-1}-1}{q-1},\;\; \text{if $x>0$}\\
                      \frac{x^{1-q}-1}{1-q},\;\; \text{if $x\leq0$}.
                     \end{cases}
 \label{eq:non_ext_log}
\end{align}
It can be again trivially checked that this reduces to the standard logarithmic function of the variable $x$ in the limit $q\rightarrow1$.

Now, we will introduce the NJL model for a nonextensive statistics. In that case, only $\Omega_{\rm med}^f$ will be modified from the expression of the full thermodynamic potential given in Eq.~\ref{eq:pot_njl_mag_final}. $\Omega_{\rm med}^f$ is given in Eq.~\ref{eq:pot_njl_med}. It is evident that the logarithms in that equation will be modified to incorporate the nonextensive parameter $q$. This will be done using Eqs.~\ref{eq:non_ext_exp} and~\ref{eq:non_ext_log}.

For the analysis in the nonextensive scenarios, we assume that there will not be any modifications to the model parameters due to the nonextensive nature of the system. So, the treatment of the nonextensive parameter $q$ will be in the same footing as the temperature and chemical potential in the model. The model parameters fitted at zero temperature and chemical potential are used to evolve the system at nonzero temperature and chemical potential. The guideline for $q$ remains the same. We extend this argument further in the presence of the magnetic field and the treatment of the magnetised medium remains analogous.

The modification in Eq.~\ref{eq:pot_njl_med} can be summarised as,
\begin{align}
\Omega_{\mathrm{med},q}^f=-\frac{N_c}{2\pi}T\sum_{f,l,s}|q_f|B\int_{-\infty}^{\infty}\frac{dp_z}{(2\pi)} \left[{\rm{ln}_q}\left(1+e_q^{-(E_f(B)-{\mu})/T}\right)+{\rm{ln}_q}\left(1+e_q^{-(E_f(B)+{\mu})/T}\right) \right].
\label{eq:pot_njl_med_q}
\end{align}
In the present analysis, we stick to the zero chemical potential scenario.\footnote{It should be noted that the formulation of the nonextensive statistics is not unique when one needs to deal with both particles and antiparticles for fermions~\cite{Rozynek:2015zca}.} In that case, ensuring $e_q(x)$ to be always a real number the condition $\big(1+(1-q)x\big)^{1/(1-q)}\geq0$ has to be satisfied. Considering $q>1$ always satisfies the condition. On the other hand, $q$-values $>1$ are also phenomenologically well motivated~\cite{ALICE:2011gmo,Cleymans:2012ya}, which we elaborate in the result section.

\section{Results}
\label{sec:res}

\subsection{In the absence of a magnetic field}
To explore the properties of a nonextensive magnetised medium we begin by revisiting the scenario in the absence of an external field. To meet that purpose we employ a $2+1$-flavour NJL model. We are more interested to explore the QCD phase diagram. Thus, chiral dynamics will hold the central attention while we utilise the NJL model.

\begin{figure}[!hbt]
 \hspace{-1.65cm}
 \includegraphics[scale=0.33]{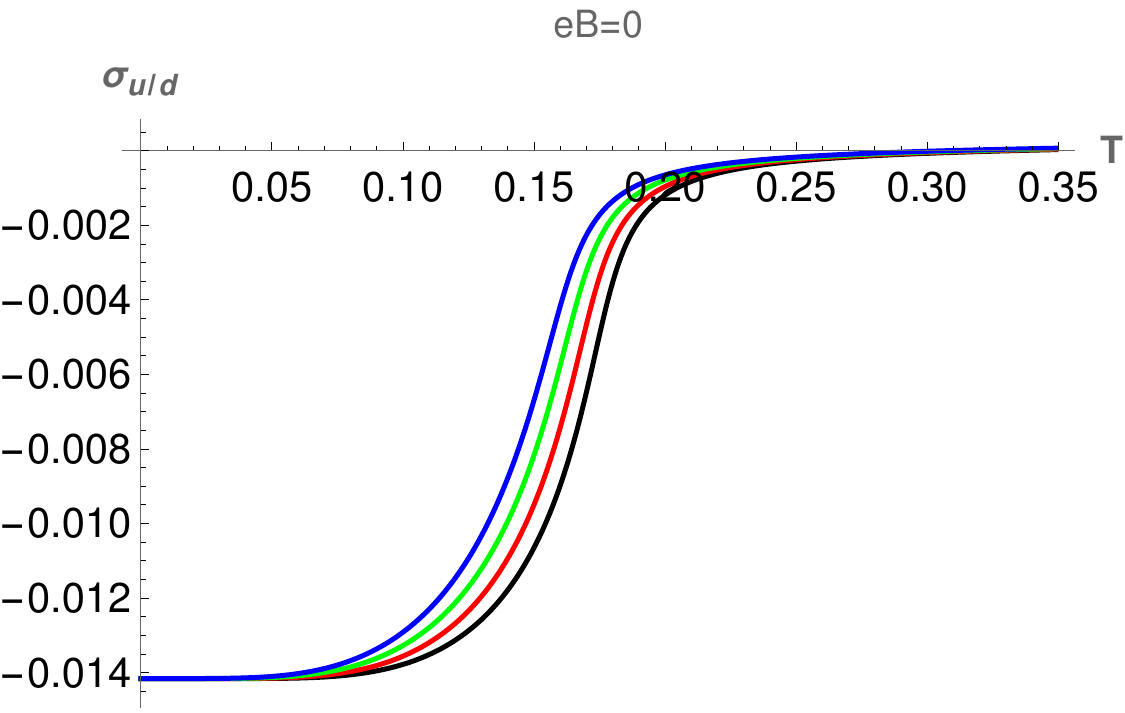}
 \hspace{0.4cm}
 \includegraphics[scale=0.33]{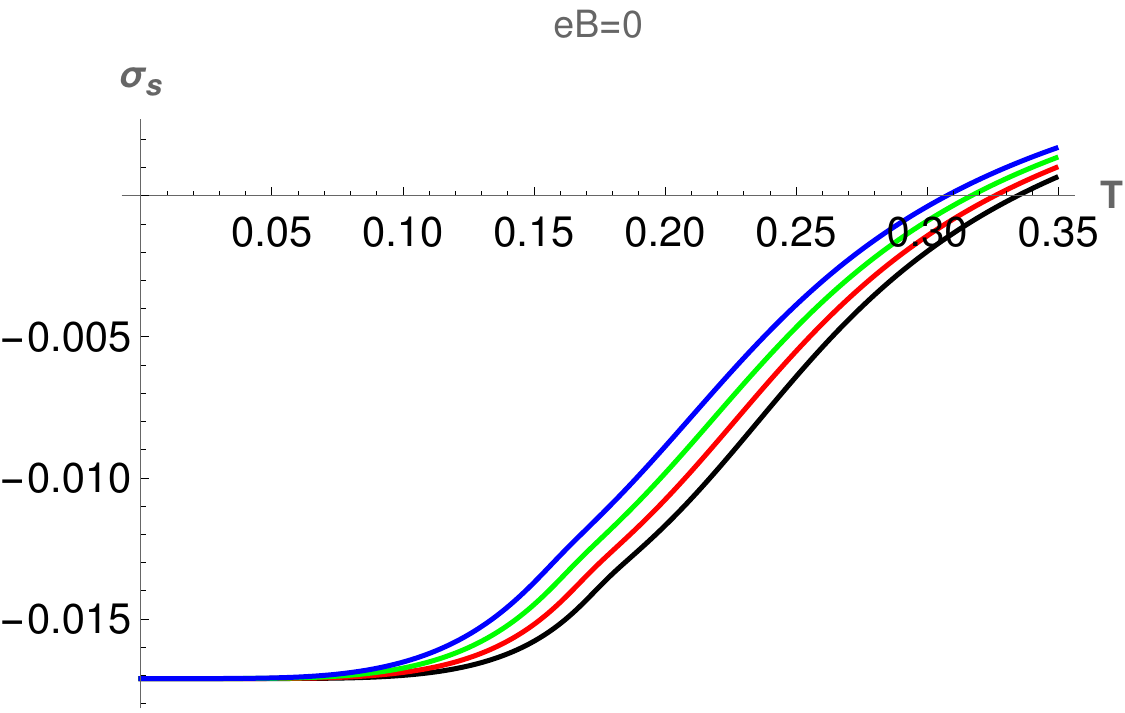}
 \includegraphics[scale=0.33]{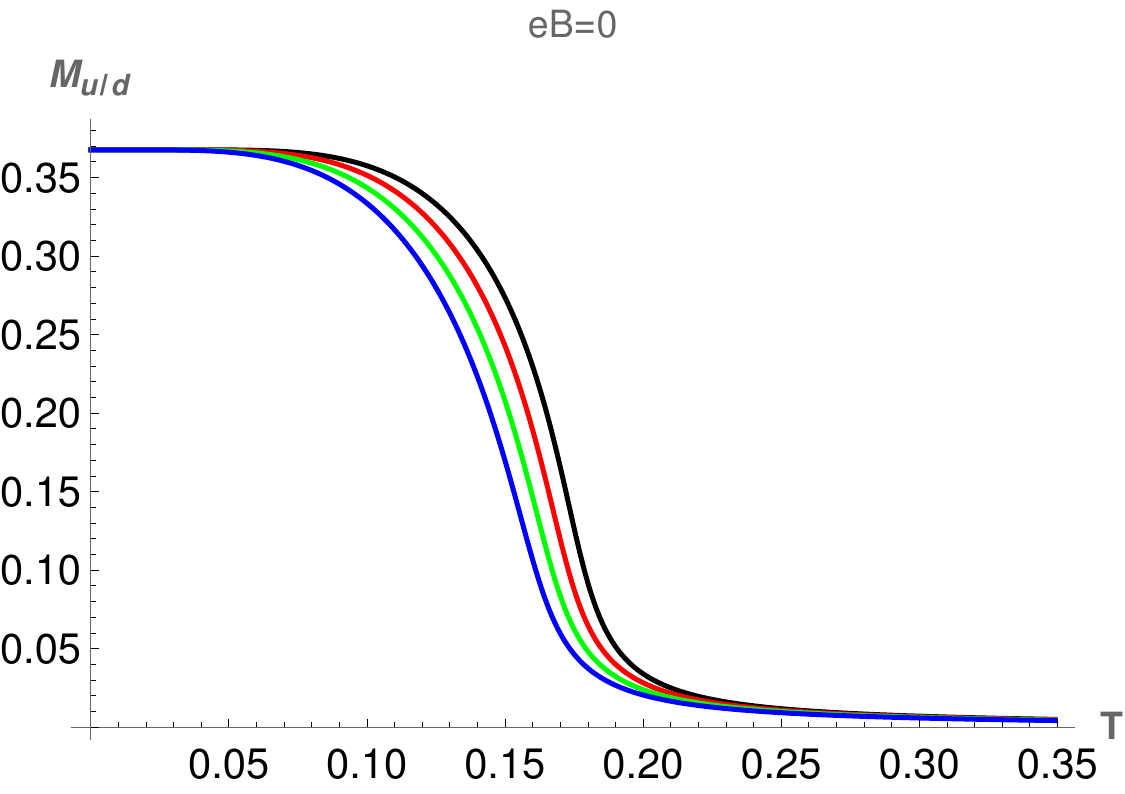}
 \hspace{0.2cm}
 \includegraphics[scale=0.32]{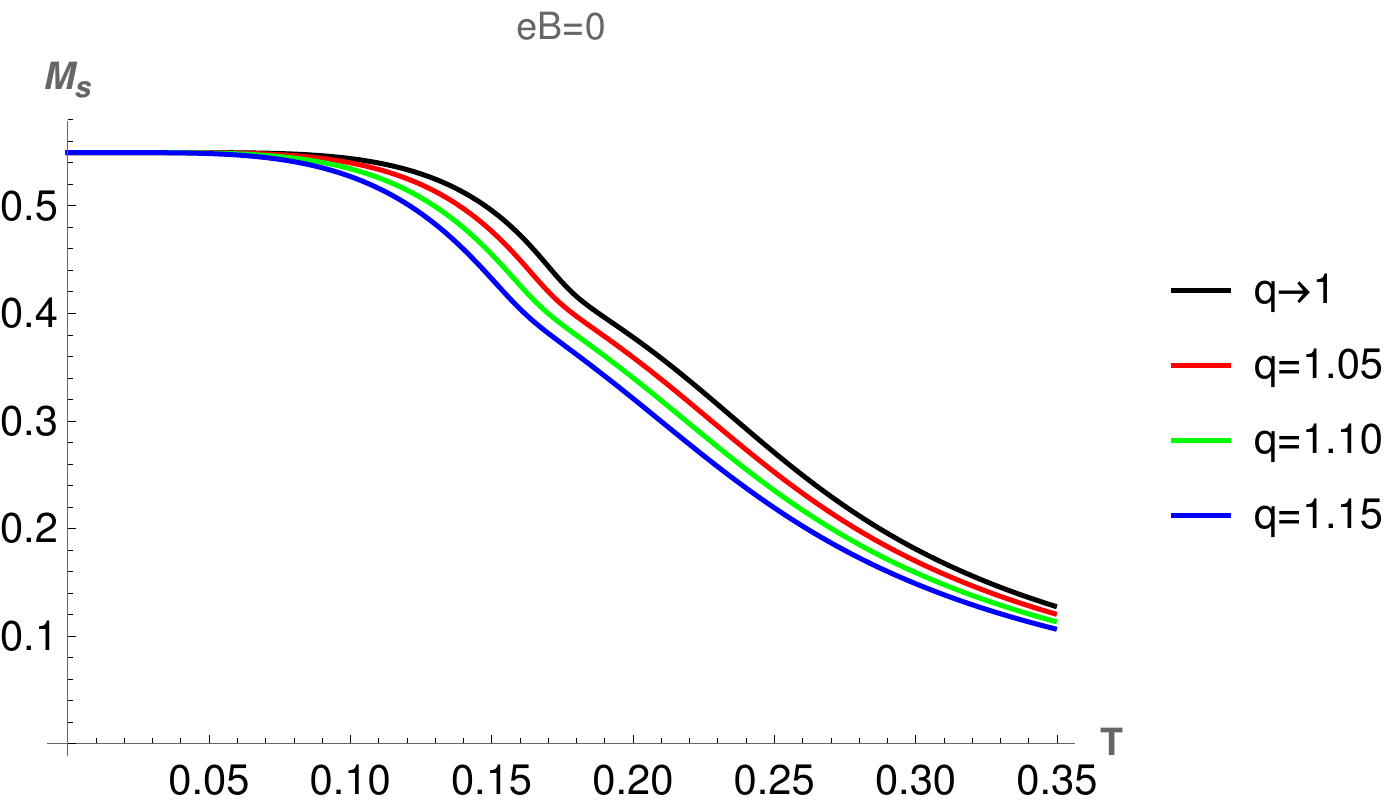}
 \caption{{\boldmath$(B=0)$}: Plot of the condensates (upper panel) and effective masses (lower panel) for different values of the $q$ parameter. Left panel: light quarks. Right panel: strange quark.}
 \label{fig:cond_n_eff_mass_eBzero}
\end{figure}
In the upper panel of Fig.~\ref{fig:cond_n_eff_mass_eBzero}, we show the plots of condensates for up/down (left panel) and strange (right panel) quarks as a function of $T$ for different values of the nonextensive parameter $q$. We observe that the chiral condensates decreases with increasing nonextensivity around the transition region. The transition also tends to be sharper for both the light and heavy quarks. Eventually, we will be more focussed on the fate of the light quarks. The plots of the strange quark condensates are displayed to keep track of the heaviest constituent of the system. To obtain the plots for the corresponding effective masses is straightforward (Eq.~\ref{eq:eff_mass}) and they reflect the same decreasing behaviour as the quark condensates with increasing $q$ values as shown in the lower panel of Fig.~\ref{fig:cond_n_eff_mass_eBzero}. Such weakening of chiral condensates and effective masses when the medium departs from its extensive nature agree with existing results.~\cite{Rozynek:2009zh,Zhao:2020xob}.

For the choice of $q>1$, in principle, one can take any higher values but in practice, it is never far away from unity, particularly in the case of HICs. In such experiments, typical values of $q$ obtained from the fits to the transverse momentum distribution for identified charged particles lies within the range of $1.1-1.2$~\cite{ALICE:2011gmo,Cleymans:2012ya}. Thus, the maximum deviation of $q$ should not exceed $20\%$. Keeping this in mind we have taken three different values of $q$, $1.05,\;1.10\;{\rm and}1.15$ to illustrate the effect of ``nonextensiveness''; with the highest value remaining well within the expected limit.

In Table~\ref{tab:tran_temp_eBzero_ne}, we quote the transition temperatures $(T_{\rm CO})$ as a function of $q$. They are obtained by tracking the inflection points of the condensates. The quoted numbers are only for the light quarks (same for both up and down quarks). To have a pictorial idea of the decrease in chiral transition temperature, we show a plot of the $T_{\rm CO}$ as a function of $q$ with the $T_{\rm CO}$ being scaled with $T_{\rm CO}(q=1)$ in Fig.~\ref{fig:Tco_ud_eBzero_njl}. There is roughly $10$ percentage decrease in the transition temperature for the highest value of $q$ that we consider here.

\begin{minipage}{\textwidth}
    \begin{minipage}[b]{0.49\textwidth}
    \centering
        \begin{tabular}{|c|ccccc|c}
        \hline
        $q$  & $1$  & $1.05$ & $1.10$ & $1.15$  & \\ \hline
        $T_{\rm CO}$ (MeV) & $173$ & $168$  & $162$  & $156$ & \\ \hline
        \end{tabular}
        \captionof{table}{{\boldmath$(B=0)$}: Chiral transition temperature as a function of $q$.}
        \label{tab:tran_temp_eBzero_ne}
    \end{minipage}
\hfill
    \begin{minipage}[b]{0.49\textwidth}
    \centering
        \includegraphics[scale=0.35]{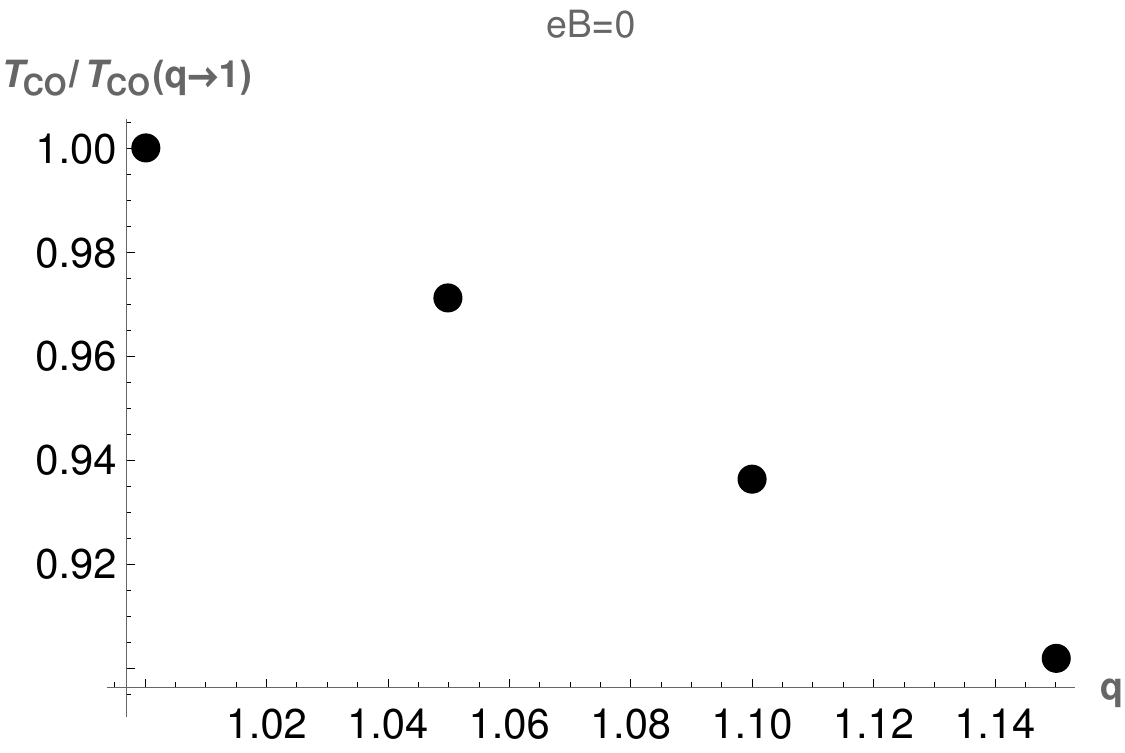}
        \captionof{figure}{{\boldmath$(B=0)$}: Scaled $T_{\rm CO}$ as a function of $q$.}
        \label{fig:Tco_ud_eBzero_njl}
    \end{minipage}
\end{minipage}

\subsection{In the presence of a magnetic field}
\label{ssec:mag_field}
In the pursuit of exploring the magnetised nonextensive medium, we proceed chronologically, i.e., we follow the steps how the NJL model in presence of the magnetic field has been developed over the years. Thus, we start with the constant $G_S$. In that case, the NJL model is known to produce the effect of MC throughout the temperature range~\cite{Gusynin:1995nb,Gusynin:1997kj}. It is interesting to check how the magnetised nonextensive medium behaves with a constant coupling.

After the discovery of the IMC effect by LQCD ~\cite{Bali:2012zg}, effective models were tweaked to reproduce that effect. One of the ways in the NJL model is to make the coupling constant dependent on $eB$~\cite{Ferreira:2014kpa} or on both $eB$ and $T$~\cite{Farias:2014eca,Farias:2016gmy}. In our present study, we choose to work with the parametrisation given in~\cite{Ferreira:2014kpa}. The parametric form solely depends on $eB$ and there is no involvement of temperature.

Since we will discuss (I)MC effects in detail, we should define them appropriately and leave no scope for confusion. The MC or IMC is always defined by looking at the behaviour of the condensates \textemdash\, its increasing value signifies catalysis and decreasing value implies to inverse catalysis. The trend of $T_{\rm CO}$ plays no role there. It is a coincidence that for the physical parameters IMC effect is accompanied with a decreasing $T_{\rm CO}$. In fact, one can conjure up scenarios in which the condensate values enhance (thus MC) but the $T_{\rm CO}$ still decreases~\cite{DElia:2018xwo}.

Apart from the NJL model~\cite{Ferreira:2014kpa,Farias:2014eca,Farias:2016gmy,Yu:2014xoa}, there are other effective models which have been utilised to study QCD in presence of a magnetic field. For example, linear sigma model (LSM) with quarks is used to explore the IMC effect in Ref.~\cite{Ayala:2014gwa}, whereas its Polyakov loop extended version (PLSM) and hadron resonance gas (HRG) models are utilised to explore thermodynamic quantities like magnetisation etc in Ref.~\cite{Tawfik:2016lih}. PLSM model has also been used further in the presence of magnetic field in Refs.~\cite{Tawfik:2016gye,Tawfik:2017cdx,Tawfik:2021eeb}.

\subsubsection{Constant coupling}
\begin{figure}[!hbt]
 \includegraphics[scale=0.27]{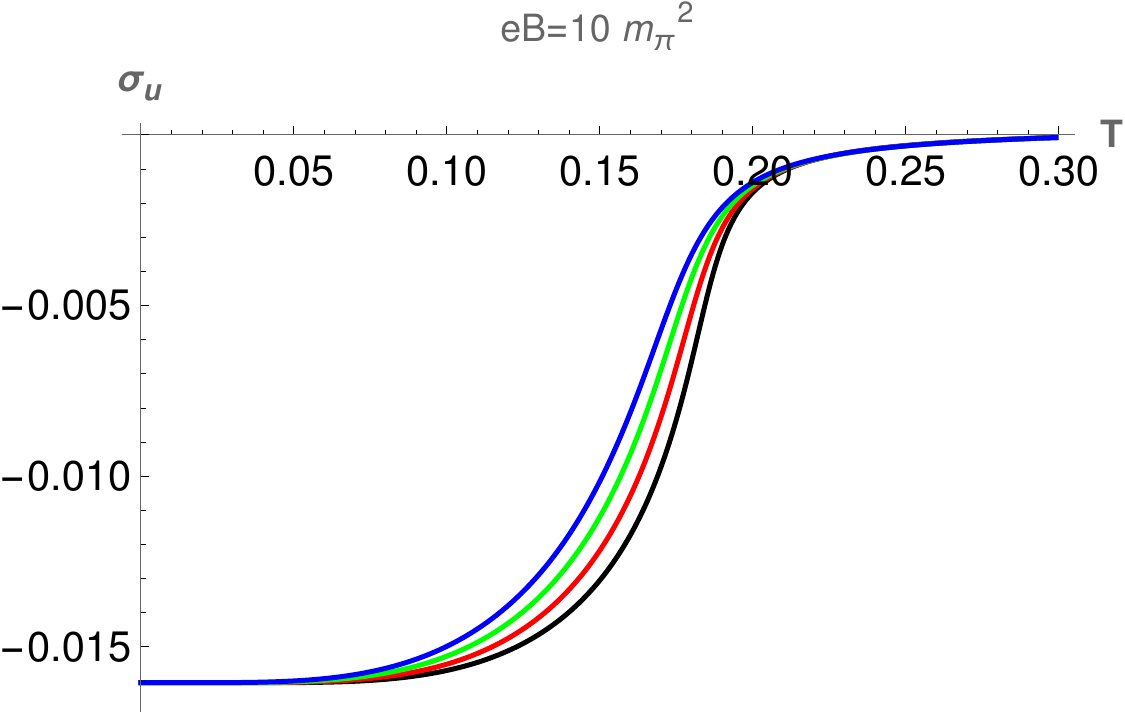}
 \includegraphics[scale=0.27]{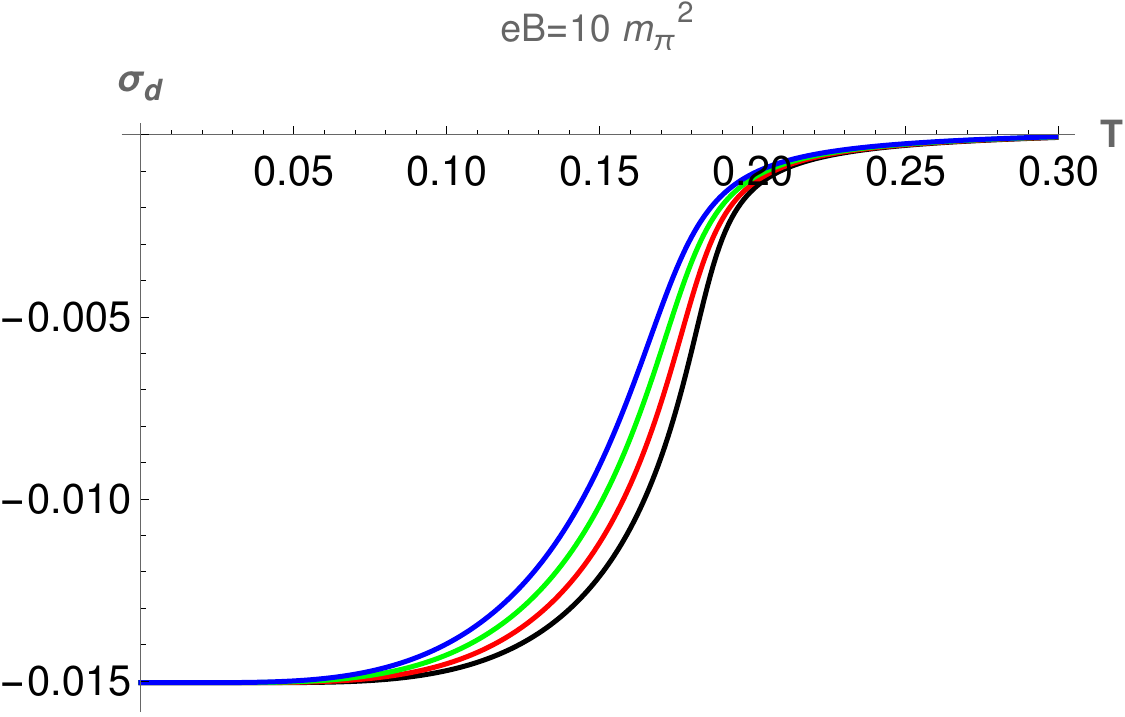}
 \includegraphics[scale=0.27]{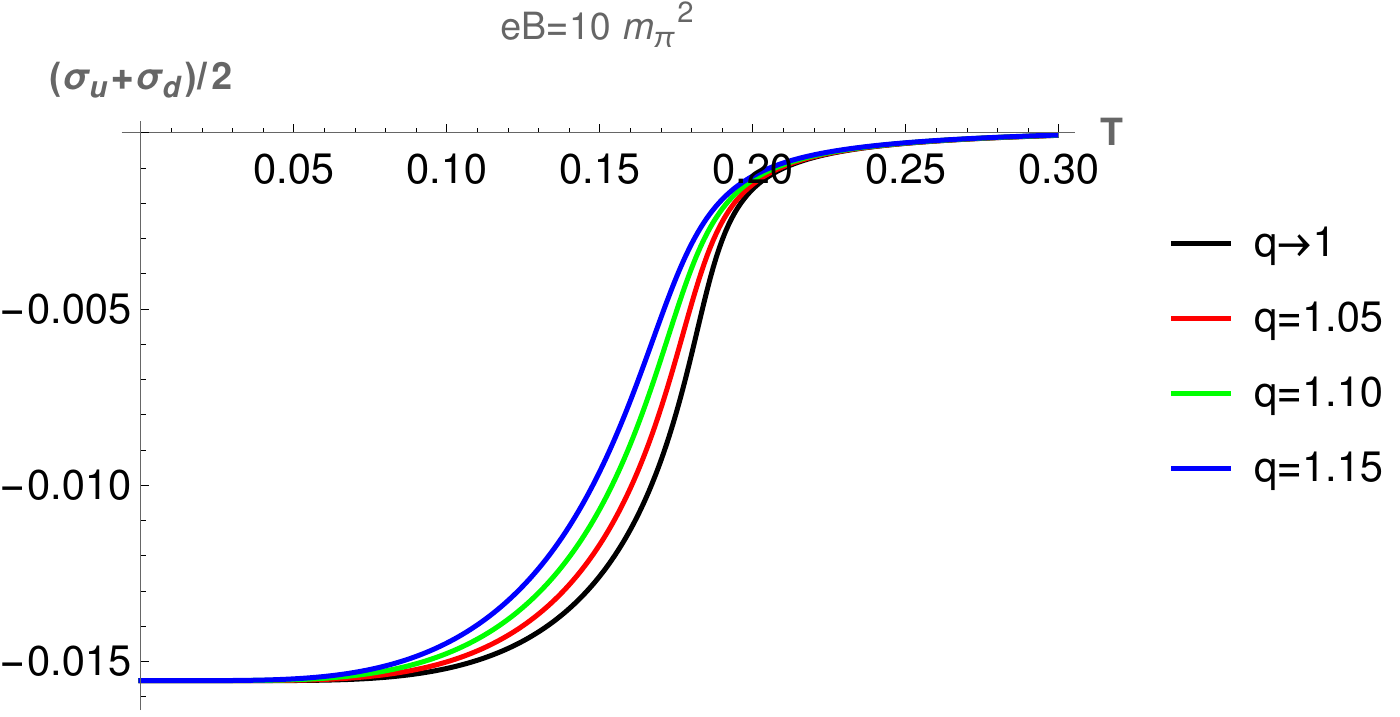}
 \caption{{\boldmath$(B\ne0,G_S^0)$}: Plot of the condensates for different values of $q$ at $eB=10\;m_\pi^2$. First two panels show up and down quark condensates, respectively. The last panel is for condensate average.}
 \label{fig:cond_ud_diff_q_fixed_eB_com}
\end{figure}

With the constant coupling, we begin the discussion by showing the plots of uniquely $(10\;m_\pi^2)$ magnetised condensates for different values of the $q$ parameter in Fig.~\ref{fig:cond_ud_diff_q_fixed_eB_com}. Because of their different coupling strength to the magnetic field the up and down quark condensates are now different. They are shown in the first and second panel. The overall effect of increasing $q$ value remains the same as it is in zero $eB$ case (Fig.~\ref{fig:cond_n_eff_mass_eBzero}), i.e., the condensate decreases with increasing $q$ for a given value of $T$. In the third panel, we describe the condensate average with the same criteria. It is important as it will be used to define the chiral crossover temperature\footnote{It is important to note that the crossover temperatures for the two light quarks may not coincide with a broken isospin symmetry.} as well as the the fate of (I)MC will be decided by looking at its behaviour. On the other hand, to track the behaviour of the individual quark condensate is important, as they respond differently in the presence of a magnetic field.

The values of the $T_{\rm CO}$ for different values of $q$ are given in Table~\ref{tab:tran_temp_ne}. As expected from the behaviour of the condensates, $T_{\rm CO}$ decreases as a function of $q$. To understand the percentage decrement, we provide with a scaled $T_{\rm CO}$ plot in Fig.~\ref{fig:Tco_ud_eB_10_njl}. The reduction in the crossover temperature can be as high as $10\%$ for the highest used value of $q$.

\begin{minipage}{\textwidth}
    \begin{minipage}[b]{0.49\textwidth}
    \centering
        \begin{tabular}{|c|ccccc|c}
        \hline
        $q$  & $1$  & $1.05$ & $1.10$ & $1.15$  & \\ \hline
        $T_{\rm CO}$ (MeV) & $182$ & $177$  & $172$  & $167$ & \\ \hline
        \end{tabular}
        \captionof{table}{{\boldmath$(G_S^0)$}: Chiral transition temperature as a function of $q$ for $eB=10\,m_\pi^2$.}
        \label{tab:tran_temp_ne}
    \end{minipage}
\hfill
    \begin{minipage}[b]{0.49\textwidth}
    \centering
        \includegraphics[scale=0.35]{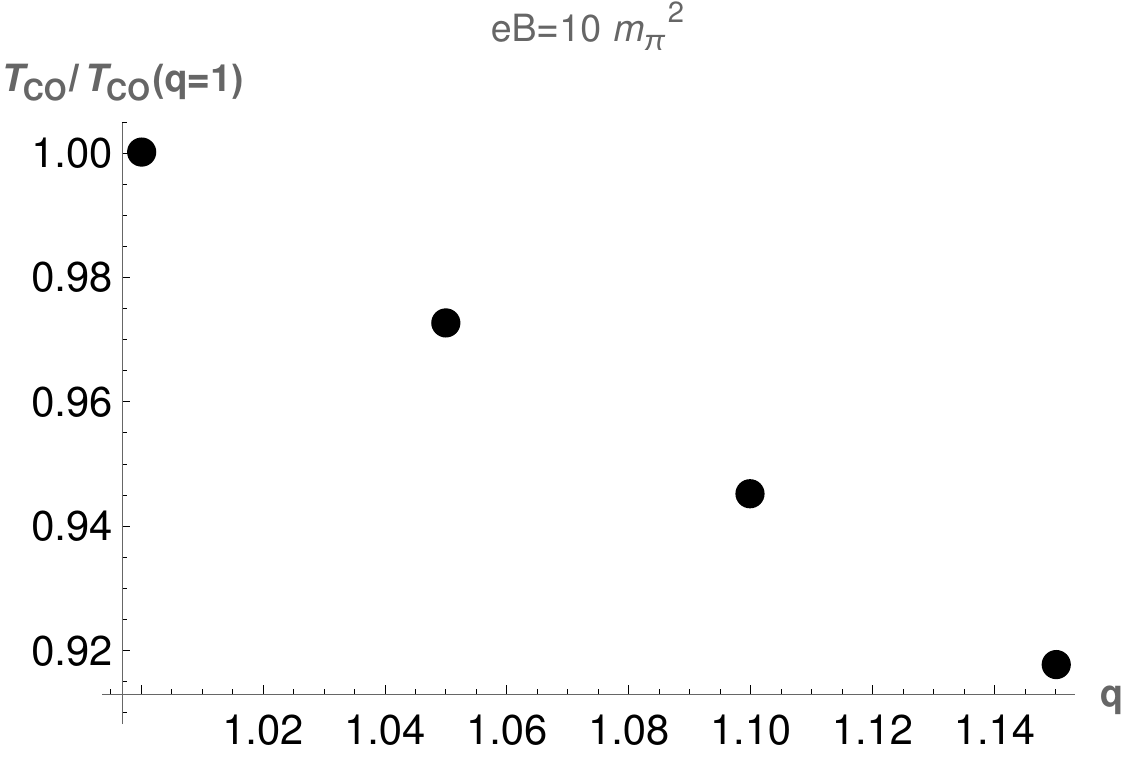}
        \captionof{figure}{{\boldmath$(G_S^0)$}: Scaled $T_{\rm CO}$ as a function of $q$ for $eB=10\,m_\pi^2$.}
        \label{fig:Tco_ud_eB_10_njl}
    \end{minipage}
\end{minipage}\\

\begin{figure}[!hbt]
 \includegraphics[scale=0.27]{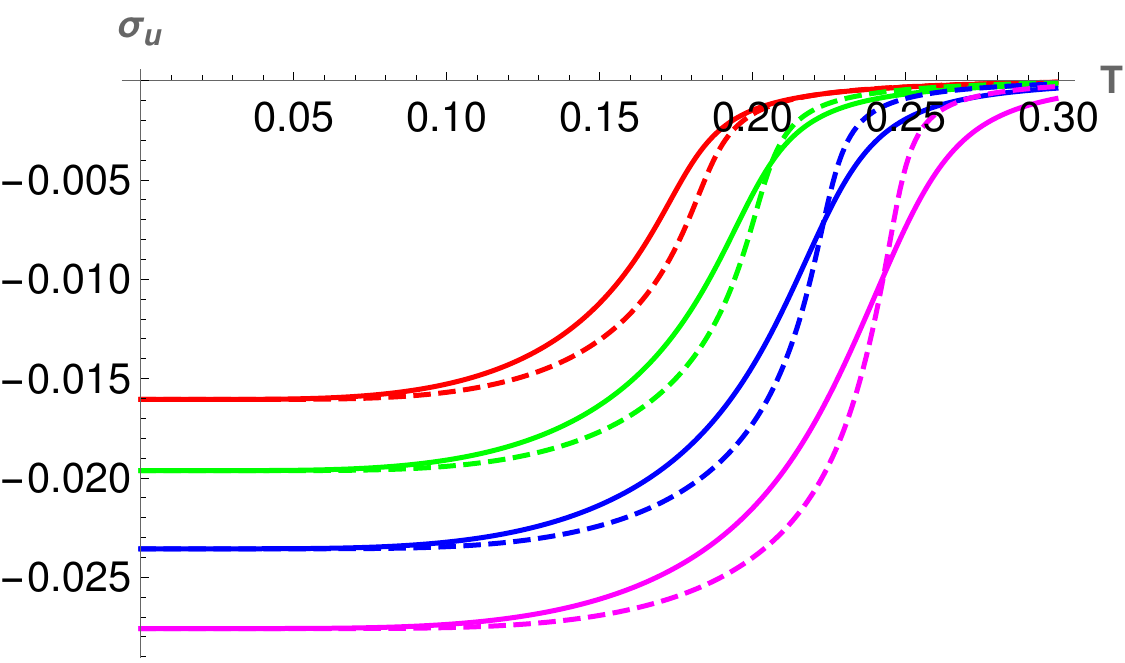}
 \includegraphics[scale=0.27]{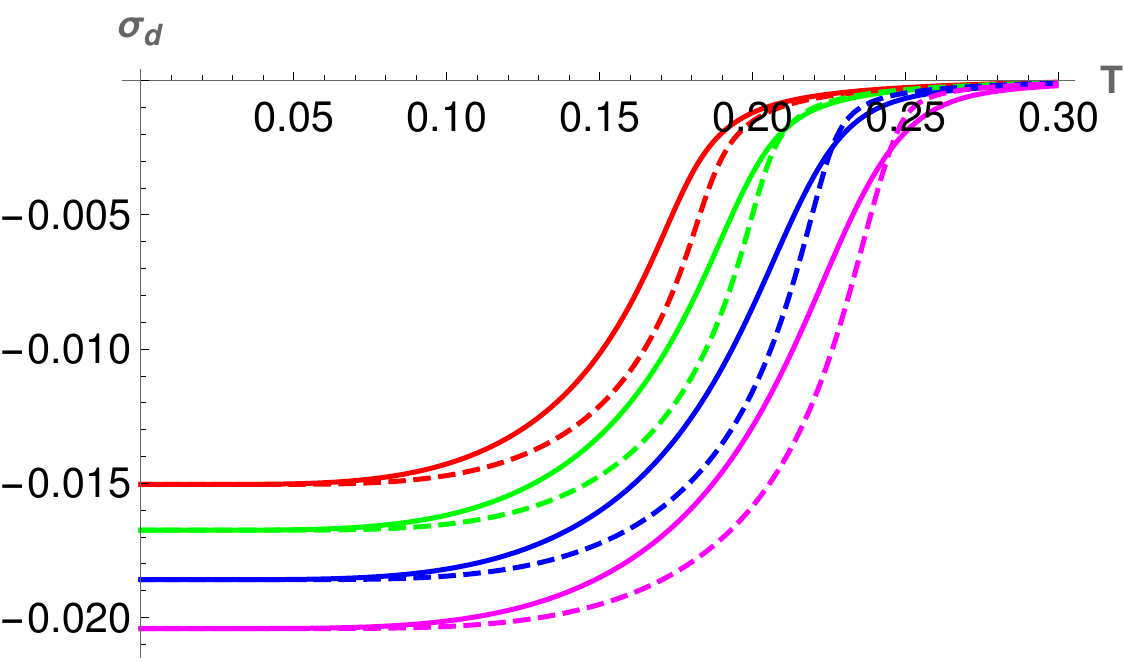}
 \includegraphics[scale=0.27]{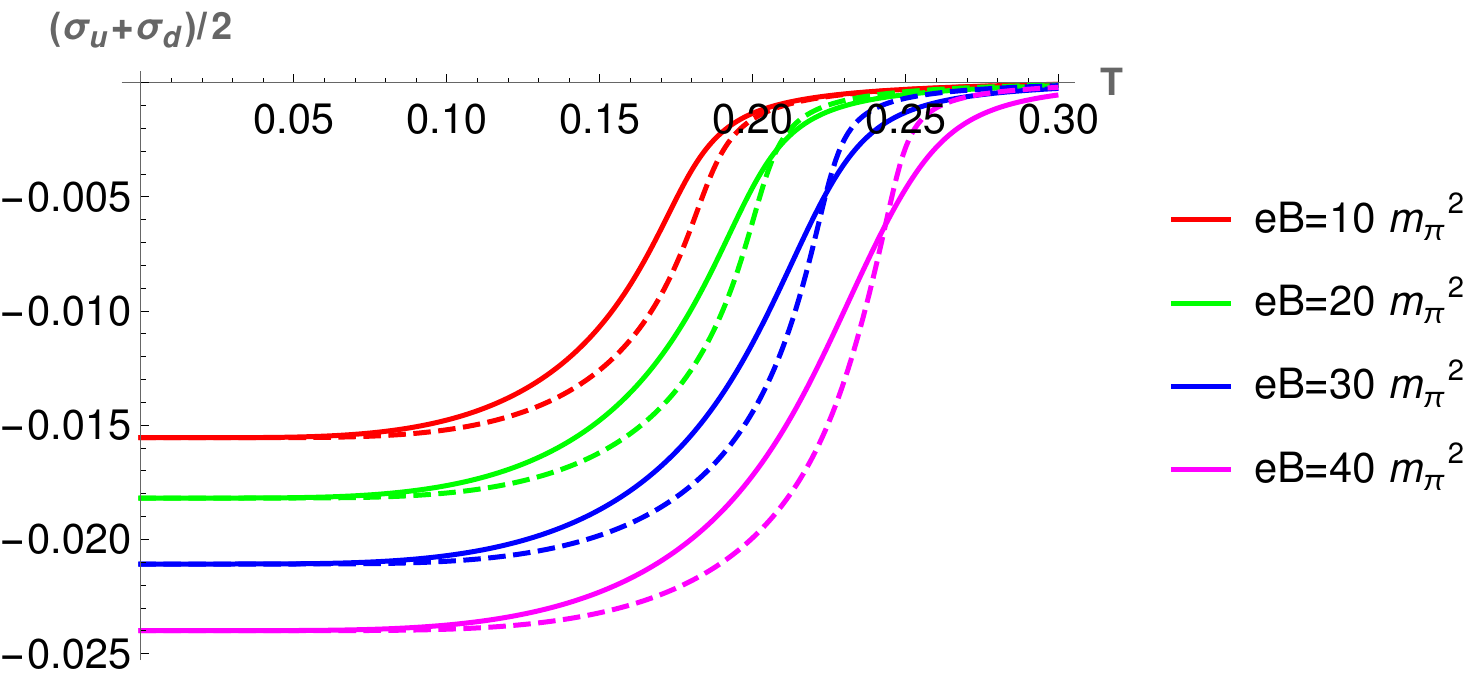}
 \caption{{\boldmath$(B\ne0,G_S^0)$}: Plot of the condensates for different values of $eB$ for both extensive $(q\rightarrow1;\; \rm{dotted\; lines})$ and nonextensive $(q=1.1;\;\rm{solid\; lines})$ cases. The three panels represent up, down and average condensates, respectively.}
 \label{fig:cond_ud_diff_eB_fixed_q_com}
\end{figure}

\vspace{0.7cm}
If the extent of the nonextensivity of the medium is fixed and the magnetic field is varied then the effects on the condensates are shown in Fig.~\ref{fig:cond_ud_diff_eB_fixed_q_com}. For the sake of comparison, the scenarios for the extensive case are also shown with the dashed lines with increasing strength of the magnetic field. The known MC effect is evident there. The MC effect also prevails in the nonextensive medium with two noticeable differences. First, at a given $eB$ value, the strength of the condensate for $q=1.1$ is smaller than the extensive case, particularly in the vicinity of the transition region. In fact, it is within this transition region that the differences between the extensive and nonextensive media are most conspicuous, especially when dealing with lower magnetic field values. Secondly, the fall of the curves are comparatively flattened in a nonextensive medium. These affect the fate of the chiral dynamics of the system which we discuss next.

\begin{minipage}{\textwidth}
    \begin{minipage}[b]{0.49\textwidth}
    \centering
        \begin{tabular}{|c|ccccccc|}
        \hline
        $eB\, (m_\pi^2)$& & $0$  & $10$ & $20$ & $30$ &  $40$ &\\ \hline
        $T_{\rm CO}$ (MeV) &$q\rightarrow1:$ & $173$ & $182$  & $200$  & $222$ & $244$ & \\
         &$q=1.10:$ & $162$ & $172$  & $192$  & $213$ & $231$ &
        \\\hline
        \end{tabular}
        \captionof{table}{{\boldmath$(G_S^0)$}: Chiral transition temperature for different values of $eB$ and $q$.}
        \label{tab:tran_temp_ne_eB}
    \end{minipage}
\hfill
    \begin{minipage}[b]{0.49\textwidth}
    \centering
        \includegraphics[scale=0.35]{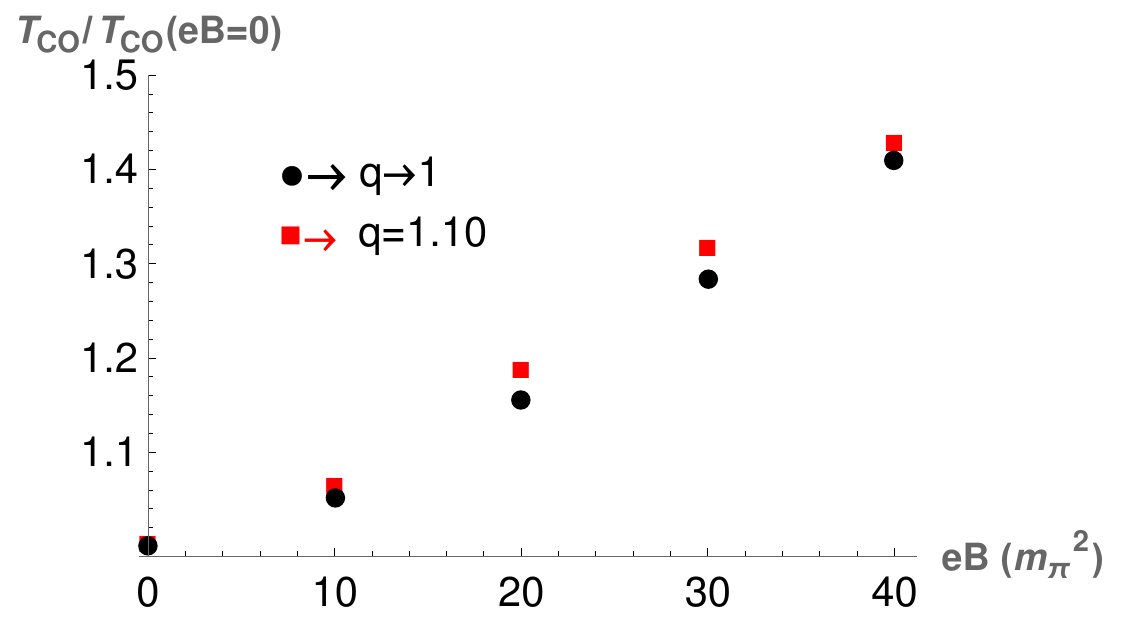}
        \captionof{figure}{{\boldmath$(G_S^0)$}: Scaled $T_{\rm CO}$ as a function of $eB$.}
        \label{fig:Tco_ud_eB_diff_q_com_njl}
    \end{minipage}
\end{minipage}\\

\vspace{0.7cm}
With MC prevailing for both extensive and nonextensive cases, the weakening of the chiral condensates and the corresponding transition temperatures are evident in Fig.~\ref{fig:cond_ud_diff_eB_fixed_q_com}. This is what reflects in the Table~\ref{tab:tran_temp_ne_eB}, where we provide $T_{\rm CO}$'s for both $q\rightarrow1$ and $q=1.10$ with different values of $eB$. Though the $T_{\rm CO}$ for a given value of $eB$ is smaller for a nonextensive medium as compared to the extensive one, the percentage change is always higher for the former for all values of $eB$ considered here. It is depicted in Fig.~\ref{fig:Tco_ud_eB_diff_q_com_njl}, where the $T_{\rm CO}$'s are scaled with the respective values at zero $eB$.

\subsubsection{Magnetic field dependent coupling constant}
For the field dependent coupling constant, we first show the condensate plots for the light quarks along with the average for the known extensive scenario. It is important to note that we are not using the modified definition of the condensates utilised in Ref.~\cite{Ferreira:2014kpa} (Eq.14) which are taken from LQCD~\cite{Bali:2012zg}. Such definitions are particularly used in effective models to have results on the same footing and thus comparable with that in LQCD.

In the present study, we do not have such obligation as we are not comparing with any existing data. Rather, we are interested to investigate the fate of the (I)MC effects for a nonextensive medium in the model itself. Thus for us, knowing the condensates and their average will be adequate to decide on the fate of the (I)MC effects and the trend of $T_{\rm CO}$. To determine the nature of the effect (whether MC or IMC) the condensate averages will be sufficient as argued in the third paragraph of Sec.~\ref{ssec:mag_field}.

Before going to discuss the results a few comments are necessary. In Ref.~\cite{Miransky:2002rp}, it was shown that the coupling constant decreases with the increasing magnetic field strength as $\alpha_s(eB)=1/\left(b\, {\rm ln}(|eB|/{\rm \Lambda_{QCD}^2})\right)$, where $\alpha_s$ is the strong interaction coupling constant and $b=(11N_c-2N_f)/6\pi$. If we consider $G_S$ to be analogous to $\alpha_s$, then it should as well decrease with $eB$. Thus, this form can be used as an ansatz to model the running coupling in NJL model. Since, we plan to make this study an extensive one, we explore this possibility as well and put it in the appendix~\ref{sec:app}.

\begin{figure}[!hbt]
 \includegraphics[scale=0.27]{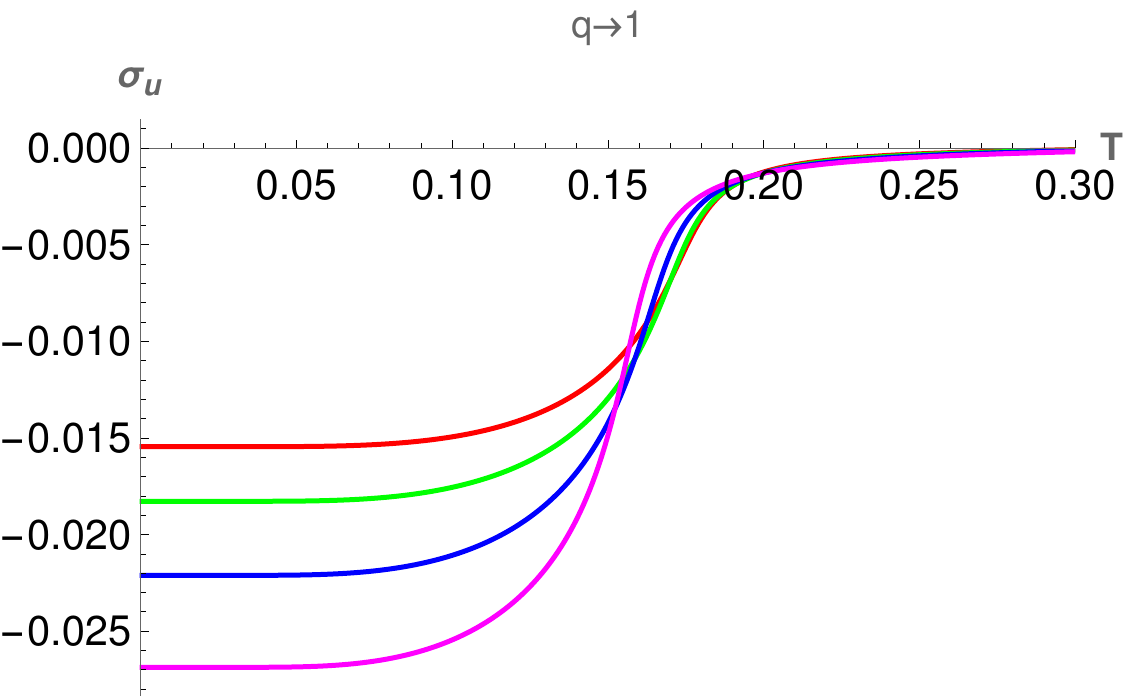}
 \includegraphics[scale=0.27]{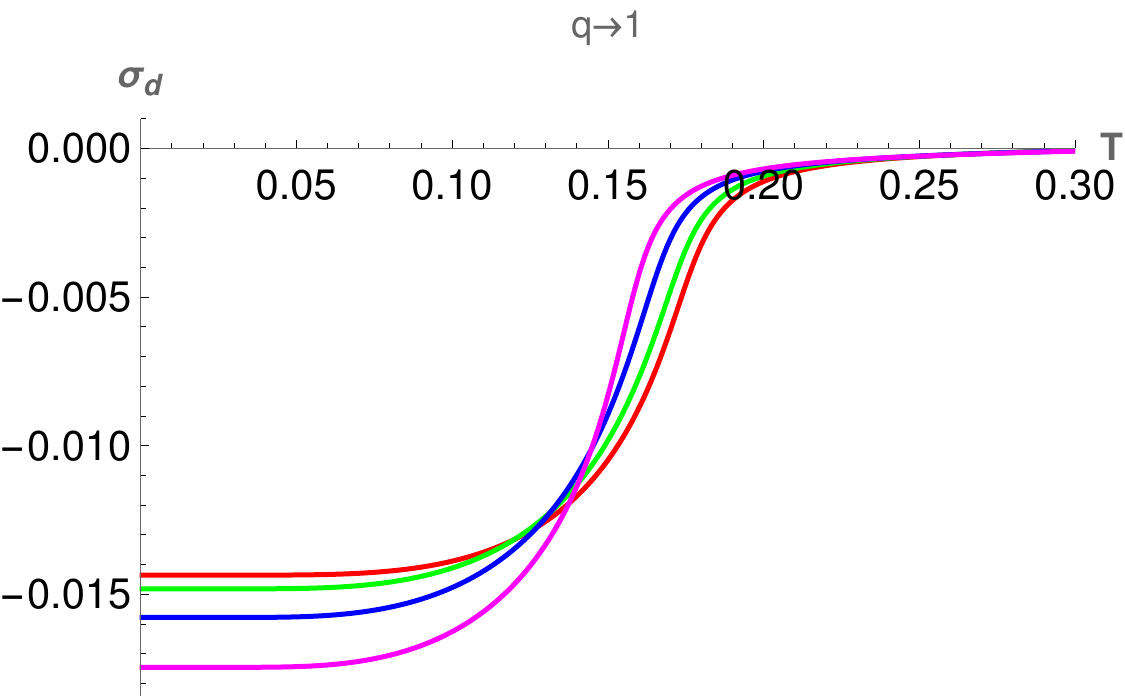}
 \includegraphics[scale=0.27]{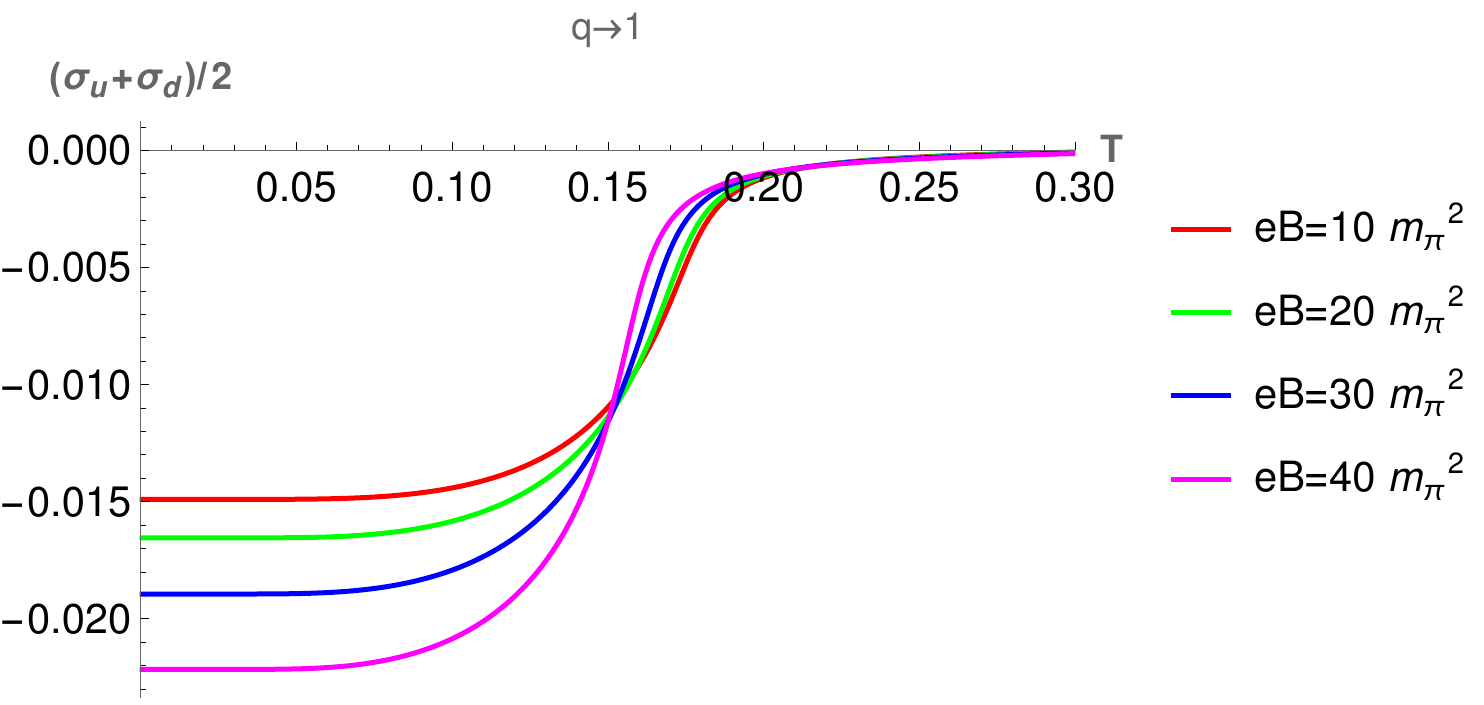}
 \caption{{\boldmath$(B\ne0,G_S(eB))$}: Plot of the light quark condensates and their average for different values of $eB$ for the extensive medium. Left to right panels represent up, down and average condensates, respectively.}
 \label{fig:cond_ud_diff_eB_GsB}
\end{figure}
In Fig.~\ref{fig:cond_ud_diff_eB_GsB}, the condensates in an extensive medium with a magnetic field dependent coupling constant are plotted. It is evident from the plots that IMC effect, expectedly, prevails around the transition region. This is similar to what is found in the existing literatures~\cite{Ferreira:2014kpa,Farias:2014eca,Farias:2016gmy,Yu:2014xoa,Farias:2014eca,Ferreira:2014kpa,Ayala:2014gwa,Farias:2016gmy,Tawfik:2016lih,Tawfik:2016gye,Tawfik:2017cdx,Tawfik:2021eeb}. Though the status of IMC will be finally decided by the trend of the condensate average, it is also interesting to observe the individual flavour. What we notice is that the up and down quarks are not affected by the IMC effect with equal magnitude. This comes as no surprise considering the different coupling strengths of up and down quarks with the magnetic field. In fact, if one calculates the crossover temperature for individual flavour, they will be different. Though, as already mentioned, the crossover temperature that we quote will always be obtained using the condensate average. It will be interesting to study how these scenarios are affected in a nonextensive medium, which we discuss next.

\begin{figure}[!hbt]
 \includegraphics[scale=0.27]{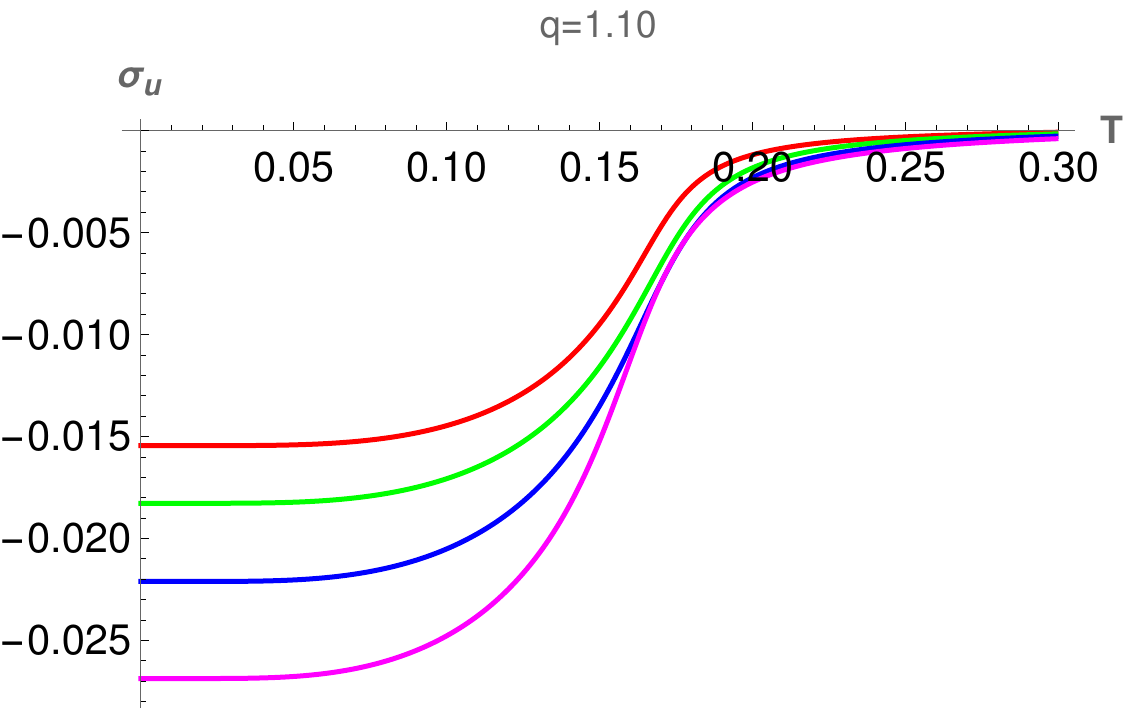}
 \includegraphics[scale=0.27]{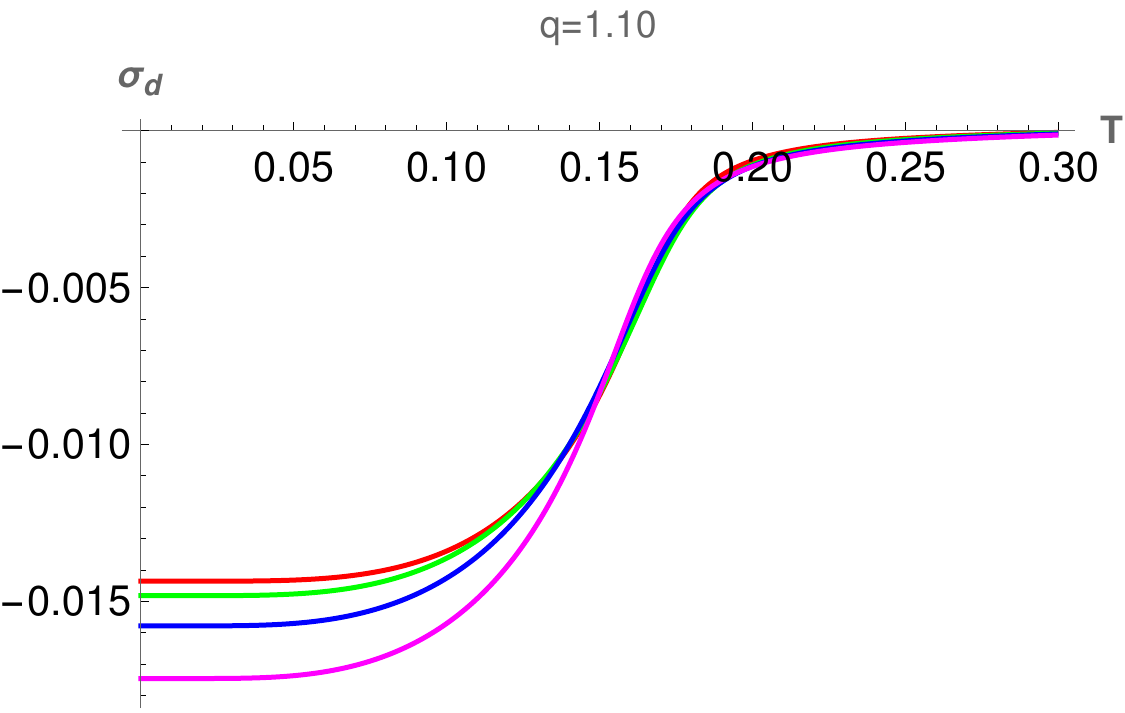}
 \includegraphics[scale=0.27]{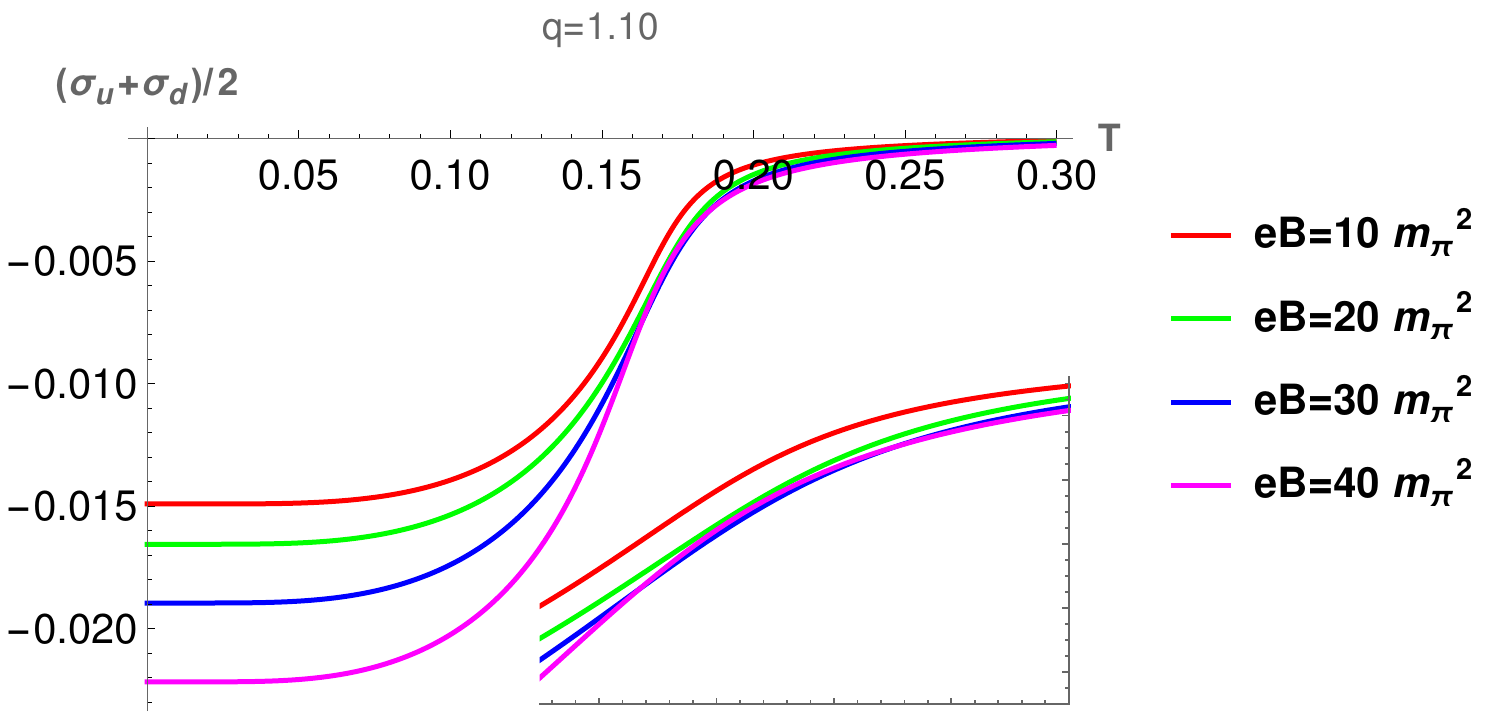}
 \caption{{\boldmath$(B\ne0,G_S(eB))$}: Plot of the condensates and their average for different values of $eB$ in a nonextensive medium. Panels are designated in the same way as in Fig.~\ref{fig:cond_ud_diff_eB_GsB}.}
 \label{fig:cond_ud_110_diff_eB}
\end{figure}
In Fig.~\ref{fig:cond_ud_110_diff_eB}, we show the same plots as those in Fig.~\ref{fig:cond_ud_diff_eB_GsB} but for a nonextensive medium with $q=1.10$. We observe some interesting changes in the behaviour of the individual as well as the average condensates. It is clear that the IMC effect for up quark is spoiled for all the strengths of the magnetic field considered here. For the down quark, the effect is present almost for all values of $eB$ but with a reduced strength. Thus, two light quarks, expectedly, behave differently in presence of a magnetic field. As for the condensate average, which we use to decide on the types of the effect in the medium, the IMC effect is ruined by the ``nonextensiveness'' of the medium except for the highest strengths of $eB$ considered here. This is elaborated with an inset plot zooming in the crossing regions. It is obvious that there is an interplay between the strengths of $q$ and $eB$, which eventually determines the types of catalysing effects (MC or IMC). We elaborate it with the next two figures. 

\begin{figure}[!hbt]
 \includegraphics[scale=0.3]{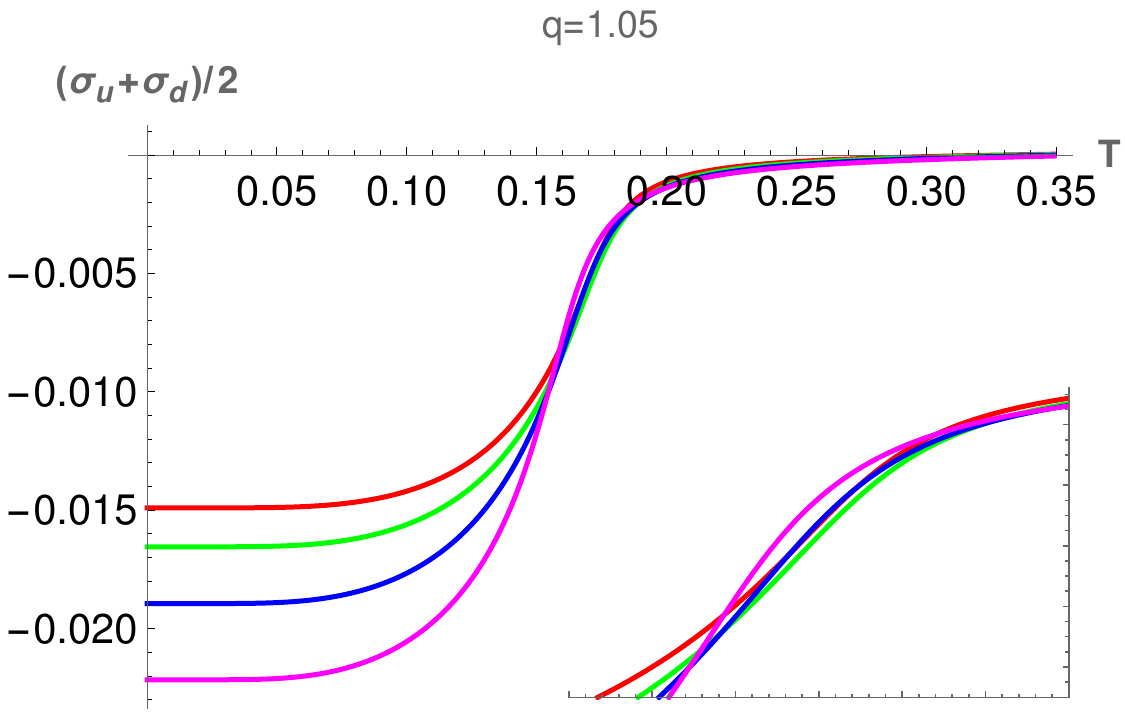}\hspace{0.5cm}
 \includegraphics[scale=0.3]{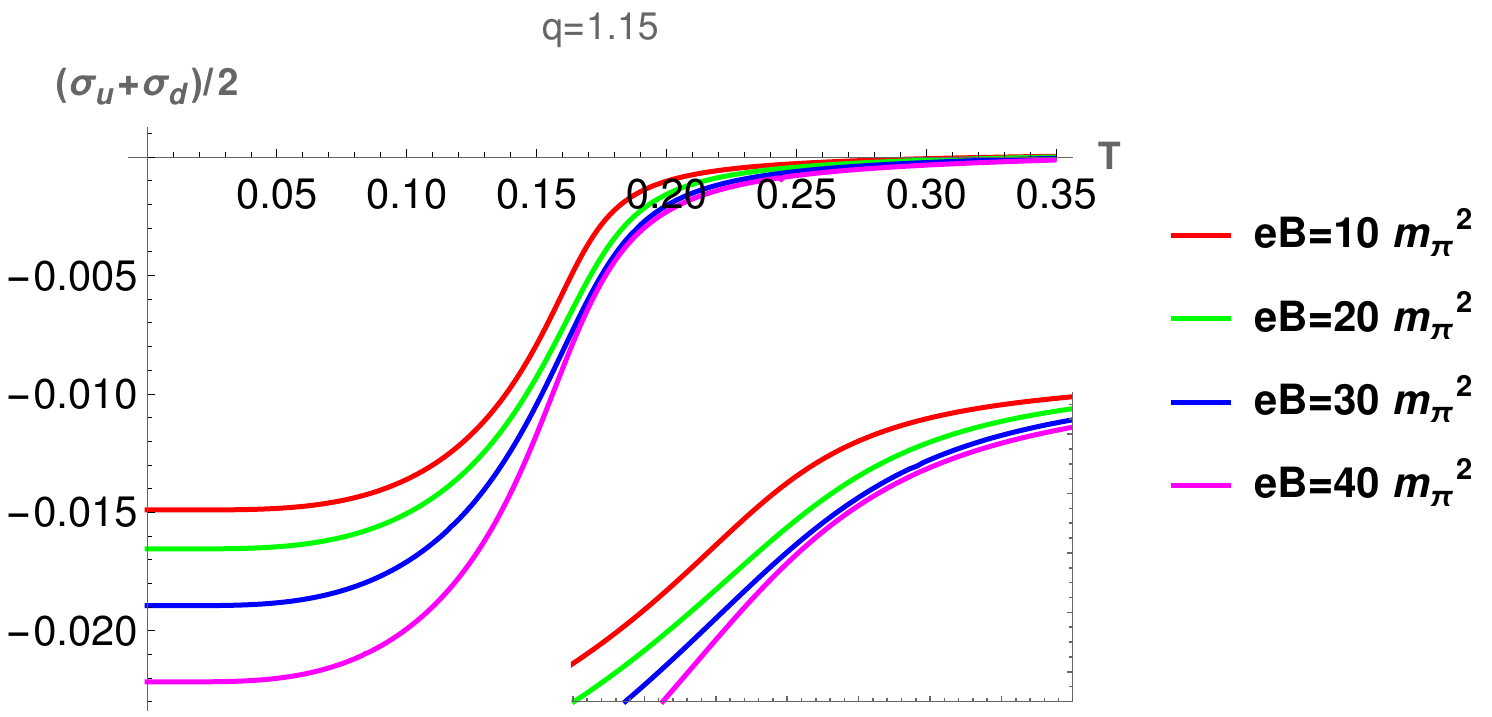}
 \caption{{\boldmath$(B\ne0,G_S(eB))$}: Plot of the condensate averages for different values of $eB$ with two different $q$'s: $q=1.05$ (left panel) and $q=1.15$ (right panel).}
 \label{fig:cond_avg_diff_eB_two_q}
\end{figure}
In Fig.~\ref{fig:cond_avg_diff_eB_two_q}, we plot the condensate averages for different values of $eB$ with two different values of $q$. These two figures help us understand the interplay between $eB$ and $q$ on determining the types of the catalysing effects in the medium. For $q=1.05$, the IMC effect reappears starting from $eB=30\,m_\pi^2$ as shown in the figure in the left panel. Thus, decreasing $q$ from $1.10$ to $1.05$ makes the IMC effect favourable for lower values of $eB$. On the other hand, if we increase $q$ to $1.15$ the IMC effect completely disappears as evident from the figure in the right panel. For both the panels, the transition regions are zoomed in for the inset plots to make the crossings of the condensates conspicuous.

Now, we make a comparison of the $T_{\rm CO}$'s for different values of $q$ including the extensive case with zero magnetic field in Table.~\ref{tab:tran_temp_ne_eB_GsB}. For $q\rightarrow1$, the decreasing nature of the crossover temperature is captured. As we crank up the nonextensive nature of the medium by increasing $q$, $T_{\rm CO}$ starts behaving differently. For $q=1.05$, Initially it does not change and then it decreases with increasing $eB$. With further increase in $q$ the trend becomes non-monotonic. For both $q=1.10$ and $1.15$, $T_{\rm CO}$ first increases and then decreases. With higher $q$ values the increase is greater.

All these points are well illustrated in the Fig.~\ref{fig:Tco_ud_eB_diff_q_com_GsB_njl}, where we can also read out the percentage change in the crossover temperatures and their trend following the liens which are solely intended to guide the eyes. We also observe that decreasing $T_{\rm CO}$ is not always accompanied with IMC. For example, at $q=1.15$ the IMC effect is absent but we still observe decreasing $T_{\rm CO}$ for the higher values of $eB$. Thus, simultaneous occurrence of the two is a mere coincidence as has been discussed in detail in Ref.~\cite{DElia:2018xwo}. There it is shown that with higher pion masses there is MC effect but with decreasing $T_{\rm CO}$.

It is also noteworthy that such non-monotonic behaviors arise in the context of an effective model due to the interplay between $eB$ and $q$, with $q \geq 1.10$. Such $q$-values render the medium non-extensive. On the other hand, all the available LQCD studies~\cite{Bali:2011qj,Endrodi:2015oba,DElia:2021yvk} demonstrate a monotonic decreasing behaviour of the crossover temperature as a function of the magnetic field, which are obviously based on extensive thermodynamics.

\begin{minipage}{\textwidth}
    \begin{minipage}[b]{0.49\textwidth}
    \centering
        \begin{tabular}{|c|ccccccc|}
        \hline
        $eB\, (m_\pi^2)$& & $0$  & $10$ & $20$ & $30$ &  $40$ &\\ \hline
        $T_{\rm CO}$ (MeV) &$q\rightarrow1:$ & $173$ & $173$  & $169$  & $163$ & $156$ &\\ 
        &$q=1.05:$ & $168$ & $168$  & $168$  & $163$ & $157$ &\\
        &$q=1.10:$ & $162$ & $164$  & $165$  & $162$ & $157$ &\\ 
        &$q=1.15:$ & $156$ & $159$  & $162$  & $160$ & $156$ &\\\hline 
        \end{tabular}
        \captionof{table}{{\boldmath$(B\ne0,G_S(eB))$}: Chiral transition temperature for different values of $eB$ and $q$.}
        \label{tab:tran_temp_ne_eB_GsB}
    \end{minipage}
\hfill
    \begin{minipage}[b]{0.49\textwidth}
    \centering
        \includegraphics[scale=0.35]{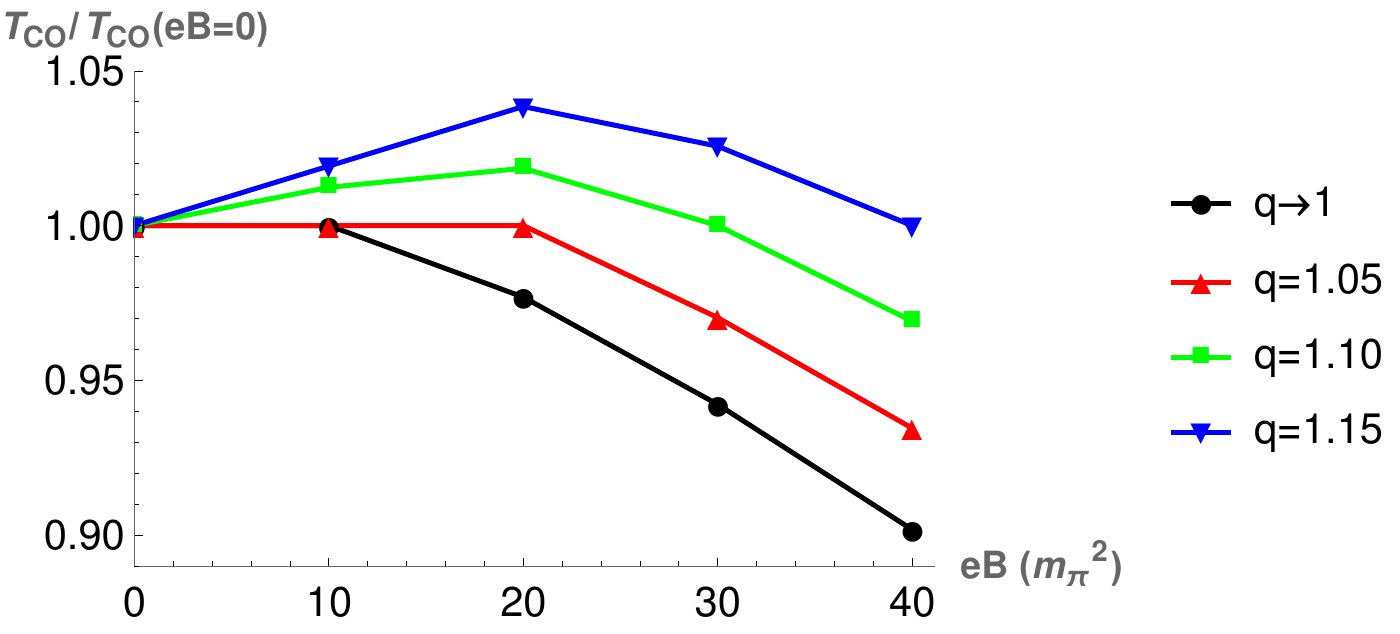}
        \captionof{figure}{{\boldmath$(B\ne0,G_S(eB))$}: Scaled $T_{\rm CO}$ as a function of $eB$.}
        \label{fig:Tco_ud_eB_diff_q_com_GsB_njl}
    \end{minipage}
\end{minipage}\\

\vspace{0.7cm}

\begin{figure}[!hbt]
 \includegraphics[scale=0.27]{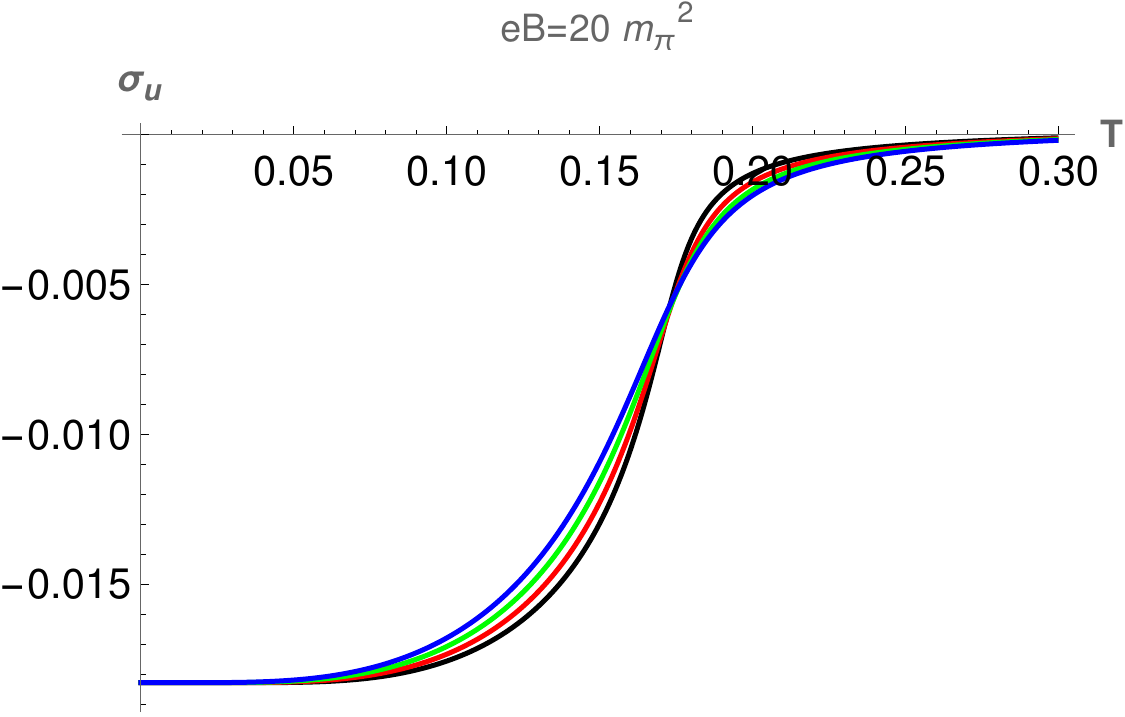}
 \includegraphics[scale=0.27]{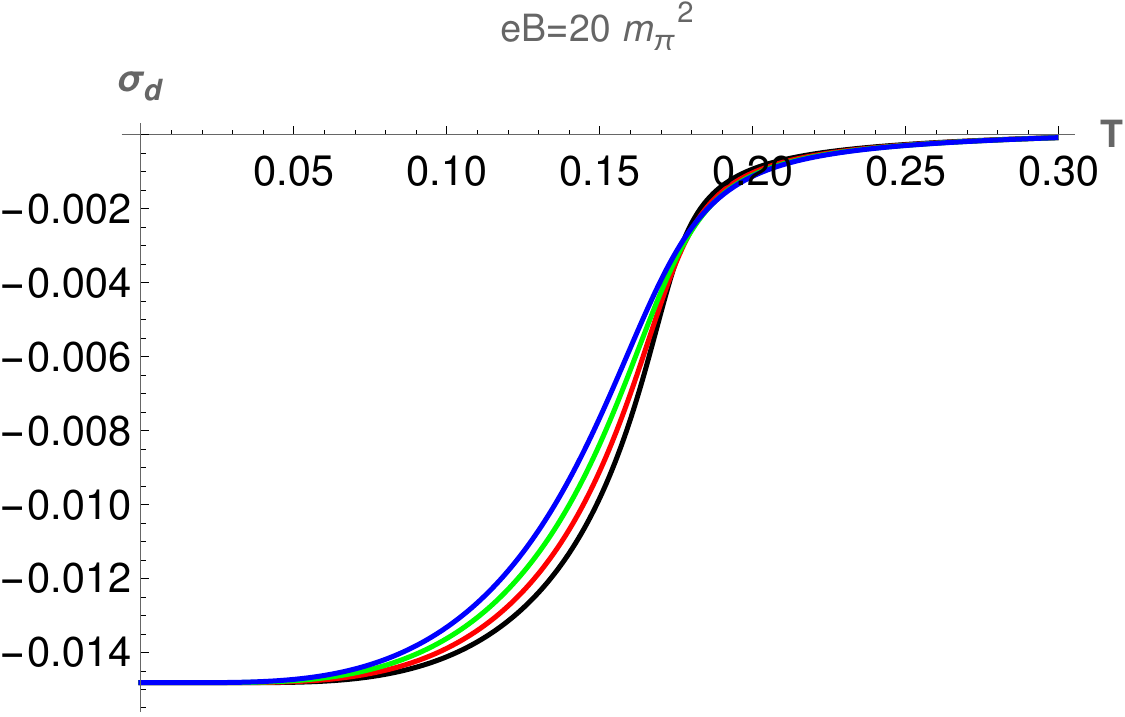}
 \includegraphics[scale=0.27]{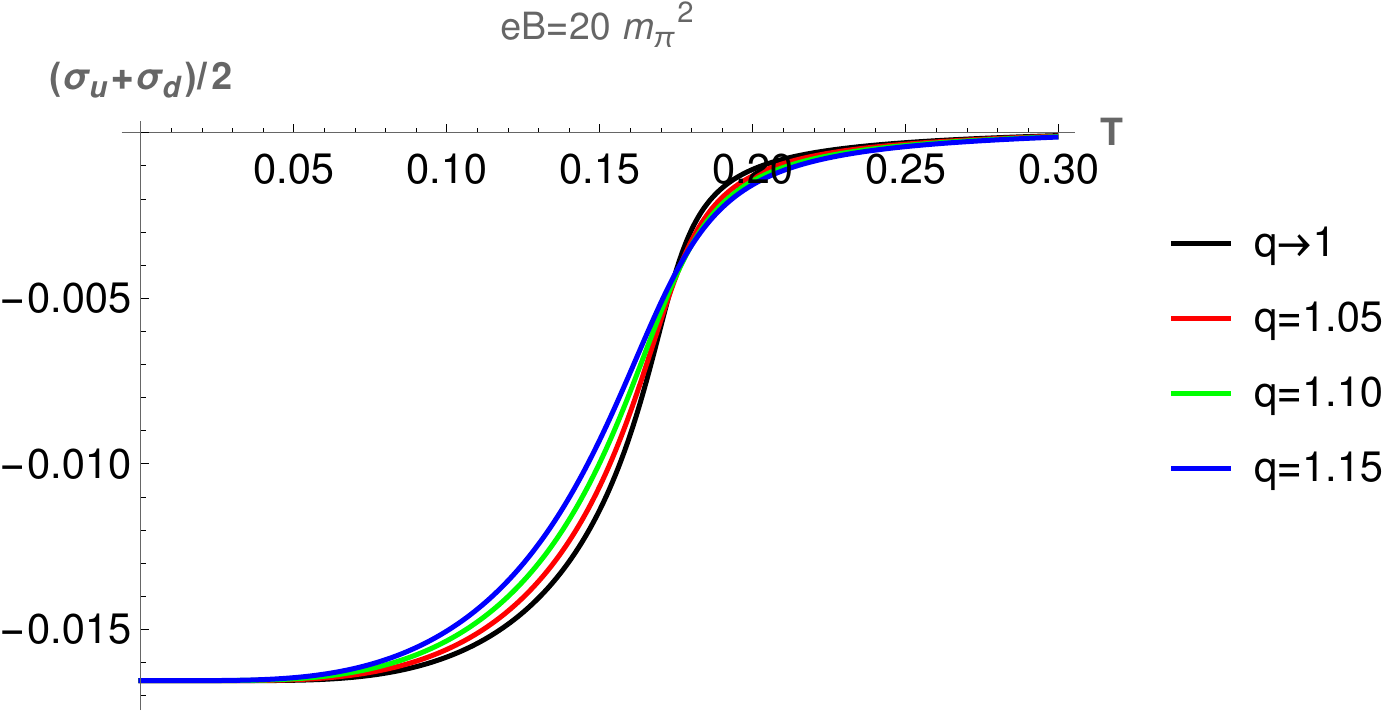}
 \caption{{\boldmath$(B\ne0,G_S(eB))$}: Plot of the condensates and their averages for different values of the $q$ parameter. Left to right panels represent up quark, down quark and average condensates, respectively.}
 \label{fig:cond_ud_diff_q_fixed_eB_GsB_com}
\end{figure}

Finally, in Fig.~\ref{fig:cond_ud_diff_q_fixed_eB_GsB_com} we show the plots for the condensates for different values of $q$ with a given $eB$. This is analogous to Fig.~\ref{fig:cond_ud_diff_q_fixed_eB_com} but for a varying coupling constant. The condensates along with their averages still display the decreasing features below the $T_{\rm CO}$'s with increasing $q$ values just like Fig.~\ref{fig:cond_ud_diff_q_fixed_eB_com} with a crossing just above the transition temperatures. This crossing appears because of the running coupling constant.

In Table.~\ref{tab:tran_temp_ne_GsB}, we quote the $T_{\rm CO}$ values for different $q$ at $eB=20\,m_\pi^2$. It is not affected by the crossings of the condensates and decreases with increasing $q$ as in the case with a constant coupling. In Fig.~\ref{fig:Tco_ud_eB_20_GsB_njl}, the scaled transition temperature is shown as a function of $q$ to elucidate the percentage change.\\

\begin{minipage}{\textwidth}
    \begin{minipage}[b]{0.49\textwidth}
    \centering
        \begin{tabular}{|c|ccccc|c}
        \hline
        $q$  & $1$  & $1.05$ & $1.10$ & $1.15$  & \\ \hline
        $T_{\rm CO}$ (MeV) & $169$ & $168$  & $165$  & $162$ & \\ \hline 
        \end{tabular}
        \captionof{table}{{\boldmath$(B\ne0,G_S(eB))$}: Chiral transition temperature as a function of $q$ for $eB=20\,m_\pi^2$.}
        \label{tab:tran_temp_ne_GsB}
    \end{minipage}
\hfill
    \begin{minipage}[b]{0.49\textwidth}
    \centering
        \includegraphics[scale=0.35]{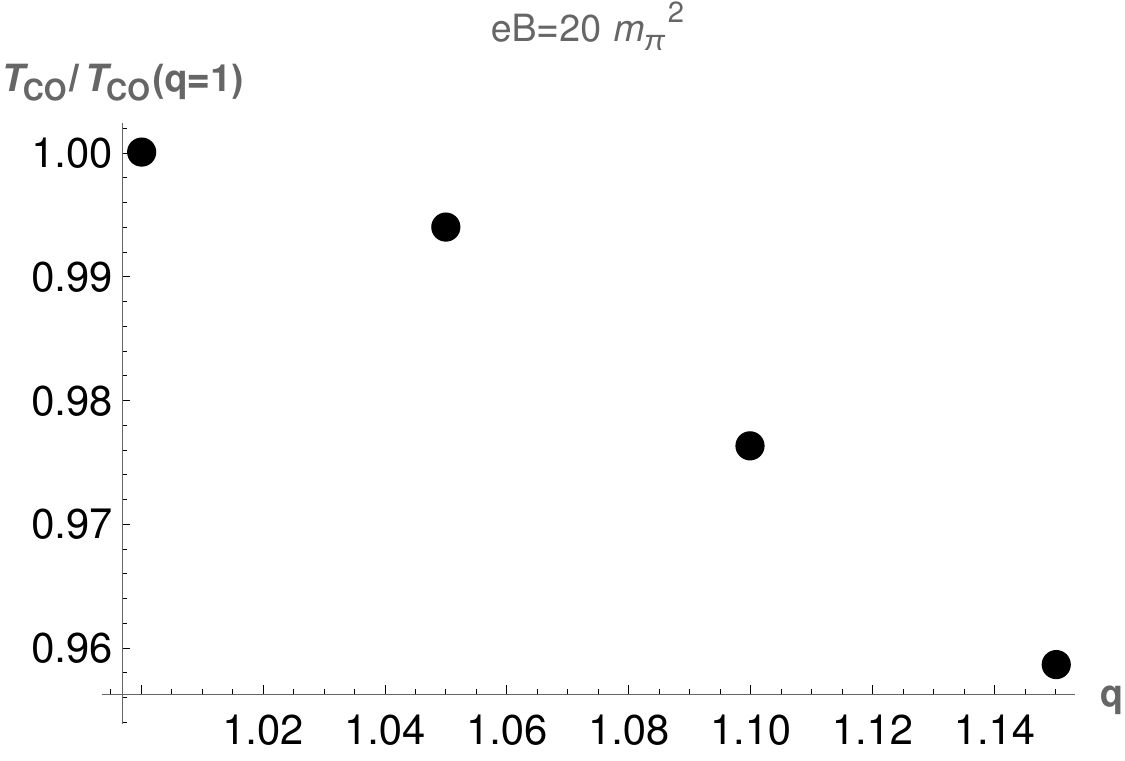}
        \captionof{figure}{{\boldmath$(B\ne0,G_S(eB))$}: Scaled $T_{\rm CO}$ as a function of $q$ for $eB=20\,m_\pi^2$.}
        \label{fig:Tco_ud_eB_20_GsB_njl}
    \end{minipage}
\end{minipage}\\

\section{Conclusion}
\label{sec:con}
We have presented an NJL model analysis of a magnetised QCD medium in a nonextensive ambience. There exist study discussing chiral symmetry restoration and other properties like meson masses in such scenarios. We add to that ongoing effort by including the magnetic field.

To achieve our goal, we utilised a $2+1$ flavour NJL model. We start the analysis by revisiting the zero magnetic field case and found that the strength of the chiral condensate decreases with increasing nonextensive parameter $q$. With increasing $q$ the chiral transition temperature decreases as well. These findings are in accord with what we already know and put our study into a perspective.

While coming to the focal point of the present work, we include the magnetic field and investigate the chiral transition for a $q$-value greater than unity. We also analysed and discussed the results for $q\rightarrow1$ wherever deemed necessary. The analysis is mainly divided into two parts \textemdash\, with i) a constant coupling $(G_S^0)$ and ii) a magnetic field dependent coupling constant $(G_S(eB))$.

For a constant coupling at a specific magnetic field strength, the impact of increasing $q$ resembles that of the zero magnetic field case, leading to a decreasing trend in both the condensates and the transition temperatures. The well-known magnetic catalysis effect persists with increasing $eB$ for a $q$-value greater than unity. However, it results in reduced strength in both the condensates and the transition temperatures compared to the extensive scenario. We also discovered that the percentage increase in the transition temperatures is consistently higher in a nonextensive scenario.

For a magnetic field dependent coupling the observations are more interesting. Depending on the strength of $q$ and $eB$ the observed IMC effect in BG statistics can be destroyed. For $q=1.10$, the IMC effect, which is nothing but the decreasing condensates with increasing strength of $eB$, is missing, except for the highest strength of magnetic field $(40 m_\pi^2)$ that we examined in this study.

There is a competition between the two parameters $q$ and $eB$ which becomes evident when we analysed for other two values of $q=1.05\; {\rm and}\; 1.15$. For $q=1.05$, the IMC effect can be observed for $eB=30m_\pi^2$, a field value lower than that required for $q=1.10$. On the other hand, at $q=1.15$, the IMC effect is totally eliminated for all the strengths of $eB$ explored here.

With $eB$-dependent $G_S$ the behaviour of the transition temperature, argued to be independent of the IMC effect, turns out to be little subtle. It also depends on both $q$ and $eB$. For $q=1.05$, it remains constant initially and then starts decreasing with increasing magnetic field. At even higher values of $q$, it exhibits a non-monotonic trend as a function of $eB$ \textemdash\, initially increasing before subsequently starts falling. The initial increment is higher for higher values of $q$.

We also explored the condensates using $G_S(eB)$ for different values of $q$ with a given $eB$. There is a crossing of the condensates at a temperature slightly higher than the transition temperature as compared to no-crossing with constant $G_S$. Also, in this case, the percentage decrease of the transition temperature is slightly lower as compared to the constant $G_S$.

However, we must keep in mind that the temperature in Tsallis statistics can have a more generalised meaning due to its non-extensive nature. In this regard, using a more generalised statistics~\cite{Hanel_2011_I,Hanel_2011_II} could be useful. Such a statistics is characterised by a unique pair of scaling exponents and the Tsallis statistics along with the Boltzmann-Gibbs statistics become special cases. It has been recently utilised in Ref.~\cite{NasserTawfik:2016sqs} to successfully reproduce the particle ratios at two different energies in HICs. There exist other works~\cite{Tawfik:2017bul,Tawfik:2018ahq} along with a review on the topic~\cite{Tawfik:2017bsy}. Thus, it will be interesting to check how much of the present conclusion gets affected in such a generalised scenario and demands a separate investigation.

Having discussed these interesting findings within a mean-field approach, it becomes intriguing to explore the effects of fluctuations by transitioning to beyond mean-field approximations. In the pursuit of robust predictions, it will be interesting to test some of the present observations using beyond mean field approaches. For example, the weakening of the chiral condensates with increasing $q$ parameter, the non-monotonic behaviour of the $T_{\rm CO}$ as a function $eB$ for $q$-values greater than unity etc. Such exercises are certainly beyond the scope of the present study and will be reported elsewhere.

As an immediate extension to the present analysis of the chiral symmetry for a nonextensive magnetised medium it would be really interesting to include the background gauge field to bring out the effect of statistical confinement. This will be particularly useful in the investigation of transport properties~\cite{Rath:2023bmx,Rath:2023abm} with effective degrees of freedom following the technique outlined in~\cite{Islam:2019tlo}. Such investigations could bear implications for HIC related observations. It also remains to be seen whether, by employing the techniques outlined in Ref.~\cite{Li:2018ygx}, the findings of the present effective model can be linked to certain phenomenological consequences. We plan to report on these in the near future.

\vspace{1cm}
{\bf Acknowledgments:} The author would like to thank Aritra Das working with whom on a different project makes him eligible to write faster and efficient codes for the present work. This research was supported in part by the ExtreMe Matter Institute EMMI at the GSI Helmholtzzentrum fuer Schwerionenforschung GmbH, Darmstadt, Germany.

\begin{appendices}
\label{sec:app}

\section{Magnetic field dependent coupling constant (simple ansatz)}
The main motivation behind the discussion in this appendix is to make the present study exhaustive and explore all possible directions that have been investigated for an extensive magnetised QCD medium using an effective model like NJL. If we follow the argument given in~\cite{Miransky:2002rp,Ferreira:2014kpa}, then $G_S$ becoming a decreasing function of $eB$ becomes an obvious fact. We exploit their argument and using the ansatz in Ref.~\cite{Ferreira:2014kpa} introduce a magnetic field dependent coupling constant
\begin{align}
 G_S(eB)=G_S^0/{\rm ln}(e+|eB|/{\rm \Lambda_{QCD}^2}).
 \label{eq:GsB_simple}
\end{align}
In case of a very high value of the magnetic field the coupling constant vanishes and bringing the magnetic field down to zero gives us back $G_S^0$.

\begin{figure}[!hbt]
 \includegraphics[scale=0.3]{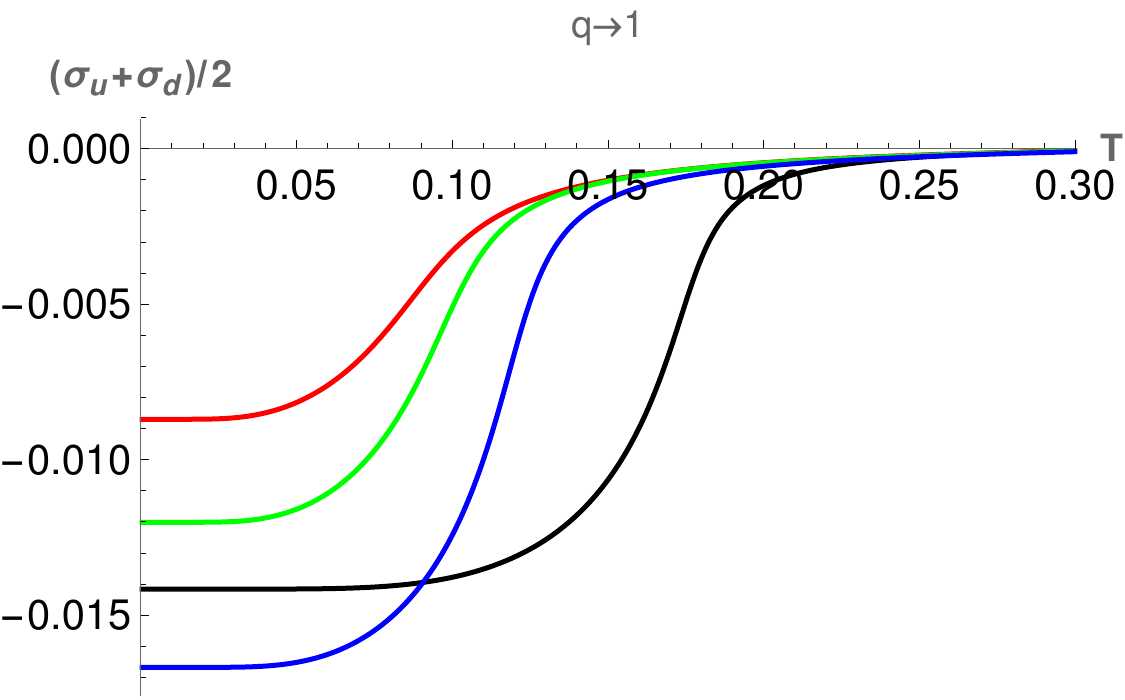}\hspace{0.5cm}
 \includegraphics[scale=0.3]{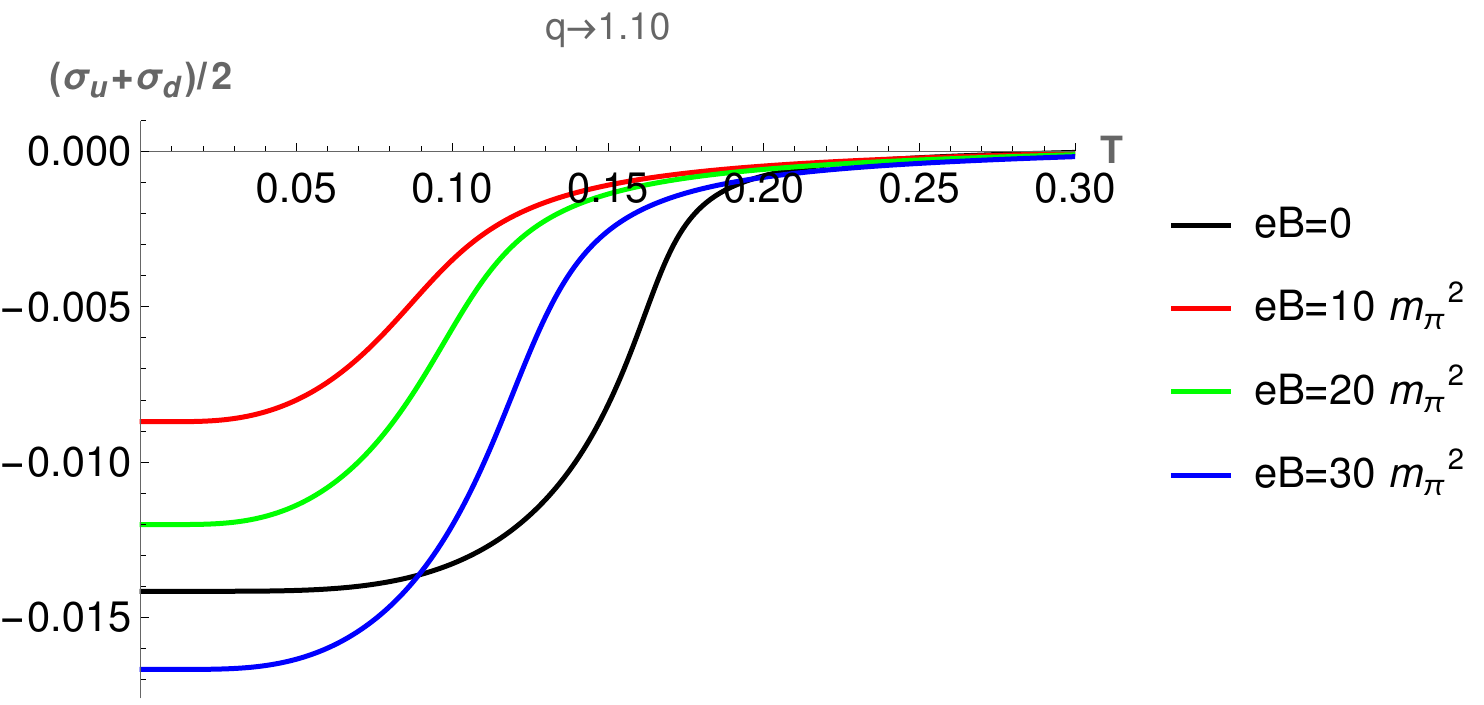}
 \caption{{\boldmath$(G_S^0/G_S(eB))$}: Plot of the condensate averages for different values of $eB$ both in the extensive (left panel) and nonextensive (right panel) scenarios.}
 \label{fig:cond_avg_diff_eB_ansatz}
\end{figure}

We explore this particular form to see its effect. We only focus on the condensate average to decide on the types of the catalysing effects and the transition temperatures. It is evident from Fig.~\ref{fig:cond_avg_diff_eB_ansatz} that with this simple ansatz (Eq.~\ref{eq:GsB_simple}) of decreasing coupling constant as a function of $eB$ one can get the IMC effect up to certain value of $eB$ albeit inconsistently. 
 
Upon concentrating on the left panel, we learn that for smaller values of magnetic field the condensate values remain always below that of zero $eB$ value. Also, the condensates with nonzero $eB$ values do not reflect the IMC effect among themselves. The crossover temperature at non-zero $eB$ values decreases initially and then starts increasing with increasing $eB$. The $T_{\rm CO}$'s, up to the values we have explored here, always remain below that in zero magnetic field case. These are in conformity with the observation in Ref~\cite{Ferreira:2014kpa}. Thus, though this simple ansatz can provide a hint of IMC effect and decreasing $T_{\rm CO}$ it fails to be in accord with the established evidence~\cite{Bali:2012zg,Bali:2011qj}. It is useful as a hint and needs to be tweaked to be successfully used in an effective model~\cite{Farias:2014eca,Farias:2016gmy,Ferreira:2014kpa}.

As we extend this analysis for a nonextensive medium $(q=1.10)$, no important change is observed and the discussion in the above paragraph remains more or less valid. The only change that takes place is the reduction in the gap of $T_{\rm CO}$'s between the zero $eB$ and highest value of $eB$ considered. Thus we can safely conclude that the magnetic field dependent coupling constant with a simple ansatz does not behave much differently in a nonextensive medium.

\end{appendices}

\bibliography{ref_common}

\begin{thebibliography}{67}%
\makeatletter
\providecommand \@ifxundefined [1]{%
 \@ifx{#1\undefined}
}%
\providecommand \@ifnum [1]{%
 \ifnum #1\expandafter \@firstoftwo
 \else \expandafter \@secondoftwo
 \fi
}%
\providecommand \@ifx [1]{%
 \ifx #1\expandafter \@firstoftwo
 \else \expandafter \@secondoftwo
 \fi
}%
\providecommand \natexlab [1]{#1}%
\providecommand \enquote  [1]{``#1''}%
\providecommand \bibnamefont  [1]{#1}%
\providecommand \bibfnamefont [1]{#1}%
\providecommand \citenamefont [1]{#1}%
\providecommand \href@noop [0]{\@secondoftwo}%
\providecommand \href [0]{\begingroup \@sanitize@url \@href}%
\providecommand \@href[1]{\@@startlink{#1}\@@href}%
\providecommand \@@href[1]{\endgroup#1\@@endlink}%
\providecommand \@sanitize@url [0]{\catcode `\\12\catcode `\$12\catcode
  `\&12\catcode `\#12\catcode `\^12\catcode `\_12\catcode `\%12\relax}%
\providecommand \@@startlink[1]{}%
\providecommand \@@endlink[0]{}%
\providecommand \url  [0]{\begingroup\@sanitize@url \@url }%
\providecommand \@url [1]{\endgroup\@href {#1}{\urlprefix }}%
\providecommand \urlprefix  [0]{URL }%
\providecommand \Eprint [0]{\href }%
\providecommand \doibase [0]{http://dx.doi.org/}%
\providecommand \selectlanguage [0]{\@gobble}%
\providecommand \bibinfo  [0]{\@secondoftwo}%
\providecommand \bibfield  [0]{\@secondoftwo}%
\providecommand \translation [1]{[#1]}%
\providecommand \BibitemOpen [0]{}%
\providecommand \bibitemStop [0]{}%
\providecommand \bibitemNoStop [0]{.\EOS\space}%
\providecommand \EOS [0]{\spacefactor3000\relax}%
\providecommand \BibitemShut  [1]{\csname bibitem#1\endcsname}%
\let\auto@bib@innerbib\@empty
\bibitem [{\citenamefont {Skokov}\ \emph {et~al.}(2009)\citenamefont {Skokov},
  \citenamefont {Illarionov},\ and\ \citenamefont {Toneev}}]{Skokov:2009qp}%
  \BibitemOpen
  \bibfield  {author} {\bibinfo {author} {\bibfnamefont {V.}~\bibnamefont
  {Skokov}}, \bibinfo {author} {\bibfnamefont {A.~Y.}\ \bibnamefont
  {Illarionov}}, \ and\ \bibinfo {author} {\bibfnamefont {V.}~\bibnamefont
  {Toneev}},\ }\href {\doibase 10.1142/S0217751X09047570} {\bibfield  {journal}
  {\bibinfo  {journal} {Int. J. Mod. Phys. A}\ }\textbf {\bibinfo {volume}
  {24}},\ \bibinfo {pages} {5925} (\bibinfo {year} {2009})},\ \Eprint
  {http://arxiv.org/abs/0907.1396} {arXiv:0907.1396 [nucl-th]} \BibitemShut
  {NoStop}%
\bibitem [{\citenamefont {Kharzeev}\ \emph {et~al.}(2013)\citenamefont
  {Kharzeev}, \citenamefont {Landsteiner}, \citenamefont {Schmitt},\ and\
  \citenamefont {Yee}}]{Kharzeev:2013jha}%
  \BibitemOpen
  \bibinfo {editor} {\bibfnamefont {D.}~\bibnamefont {Kharzeev}}, \bibinfo
  {editor} {\bibfnamefont {K.}~\bibnamefont {Landsteiner}}, \bibinfo {editor}
  {\bibfnamefont {A.}~\bibnamefont {Schmitt}}, \ and\ \bibinfo {editor}
  {\bibfnamefont {H.-U.}\ \bibnamefont {Yee}},\ eds.,\ \href {\doibase
  10.1007/978-3-642-37305-3} {\emph {\bibinfo {title} {{Strongly Interacting
  Matter in Magnetic Fields}}}},\ Vol.\ \bibinfo {volume} {871}\ (\bibinfo
  {publisher} {Springer},\ \bibinfo {year} {2013})\BibitemShut {NoStop}%
\bibitem [{\citenamefont {Miransky}\ and\ \citenamefont
  {Shovkovy}(2015)}]{Miransky:2015ava}%
  \BibitemOpen
  \bibfield  {author} {\bibinfo {author} {\bibfnamefont {V.~A.}\ \bibnamefont
  {Miransky}}\ and\ \bibinfo {author} {\bibfnamefont {I.~A.}\ \bibnamefont
  {Shovkovy}},\ }\href {\doibase 10.1016/j.physrep.2015.02.003} {\bibfield
  {journal} {\bibinfo  {journal} {Phys. Rept.}\ }\textbf {\bibinfo {volume}
  {576}},\ \bibinfo {pages} {1} (\bibinfo {year} {2015})},\ \Eprint
  {http://arxiv.org/abs/1503.00732} {arXiv:1503.00732 [hep-ph]} \BibitemShut
  {NoStop}%
\bibitem [{\citenamefont {Gusynin}\ \emph {et~al.}(1994)\citenamefont
  {Gusynin}, \citenamefont {Miransky},\ and\ \citenamefont
  {Shovkovy}}]{Gusynin:1994re}%
  \BibitemOpen
  \bibfield  {author} {\bibinfo {author} {\bibfnamefont {V.~P.}\ \bibnamefont
  {Gusynin}}, \bibinfo {author} {\bibfnamefont {V.~A.}\ \bibnamefont
  {Miransky}}, \ and\ \bibinfo {author} {\bibfnamefont {I.~A.}\ \bibnamefont
  {Shovkovy}},\ }\href {\doibase 10.1103/PhysRevLett.73.3499} {\bibfield
  {journal} {\bibinfo  {journal} {Phys. Rev. Lett.}\ }\textbf {\bibinfo
  {volume} {73}},\ \bibinfo {pages} {3499} (\bibinfo {year} {1994})},\ \bibinfo
  {note} {[Erratum: Phys.Rev.Lett. 76, 1005 (1996)]},\ \Eprint
  {http://arxiv.org/abs/hep-ph/9405262} {arXiv:hep-ph/9405262} \BibitemShut
  {NoStop}%
\bibitem [{\citenamefont {Bali}\ \emph
  {et~al.}(2012{\natexlab{a}})\citenamefont {Bali}, \citenamefont {Bruckmann},
  \citenamefont {Endrodi}, \citenamefont {Fodor}, \citenamefont {Katz},\ and\
  \citenamefont {Schafer}}]{Bali:2012zg}%
  \BibitemOpen
  \bibfield  {author} {\bibinfo {author} {\bibfnamefont {G.~S.}\ \bibnamefont
  {Bali}}, \bibinfo {author} {\bibfnamefont {F.}~\bibnamefont {Bruckmann}},
  \bibinfo {author} {\bibfnamefont {G.}~\bibnamefont {Endrodi}}, \bibinfo
  {author} {\bibfnamefont {Z.}~\bibnamefont {Fodor}}, \bibinfo {author}
  {\bibfnamefont {S.~D.}\ \bibnamefont {Katz}}, \ and\ \bibinfo {author}
  {\bibfnamefont {A.}~\bibnamefont {Schafer}},\ }\href {\doibase
  10.1103/PhysRevD.86.071502} {\bibfield  {journal} {\bibinfo  {journal} {Phys.
  Rev. D}\ }\textbf {\bibinfo {volume} {86}},\ \bibinfo {pages} {071502}
  (\bibinfo {year} {2012}{\natexlab{a}})},\ \Eprint
  {http://arxiv.org/abs/1206.4205} {arXiv:1206.4205 [hep-lat]} \BibitemShut
  {NoStop}%
\bibitem [{\citenamefont {Bali}\ \emph
  {et~al.}(2012{\natexlab{b}})\citenamefont {Bali}, \citenamefont {Bruckmann},
  \citenamefont {Endrodi}, \citenamefont {Fodor}, \citenamefont {Katz},
  \citenamefont {Krieg}, \citenamefont {Schafer},\ and\ \citenamefont
  {Szabo}}]{Bali:2011qj}%
  \BibitemOpen
  \bibfield  {author} {\bibinfo {author} {\bibfnamefont {G.~S.}\ \bibnamefont
  {Bali}}, \bibinfo {author} {\bibfnamefont {F.}~\bibnamefont {Bruckmann}},
  \bibinfo {author} {\bibfnamefont {G.}~\bibnamefont {Endrodi}}, \bibinfo
  {author} {\bibfnamefont {Z.}~\bibnamefont {Fodor}}, \bibinfo {author}
  {\bibfnamefont {S.~D.}\ \bibnamefont {Katz}}, \bibinfo {author}
  {\bibfnamefont {S.}~\bibnamefont {Krieg}}, \bibinfo {author} {\bibfnamefont
  {A.}~\bibnamefont {Schafer}}, \ and\ \bibinfo {author} {\bibfnamefont
  {K.~K.}\ \bibnamefont {Szabo}},\ }\href {\doibase 10.1007/JHEP02(2012)044}
  {\bibfield  {journal} {\bibinfo  {journal} {JHEP}\ }\textbf {\bibinfo
  {volume} {02}},\ \bibinfo {pages} {044} (\bibinfo {year}
  {2012}{\natexlab{b}})},\ \Eprint {http://arxiv.org/abs/1111.4956}
  {arXiv:1111.4956 [hep-lat]} \BibitemShut {NoStop}%
\bibitem [{\citenamefont {Fukushima}\ \emph {et~al.}(2008)\citenamefont
  {Fukushima}, \citenamefont {Kharzeev},\ and\ \citenamefont
  {Warringa}}]{Fukushima:2008xe}%
  \BibitemOpen
  \bibfield  {author} {\bibinfo {author} {\bibfnamefont {K.}~\bibnamefont
  {Fukushima}}, \bibinfo {author} {\bibfnamefont {D.~E.}\ \bibnamefont
  {Kharzeev}}, \ and\ \bibinfo {author} {\bibfnamefont {H.~J.}\ \bibnamefont
  {Warringa}},\ }\href {\doibase 10.1103/PhysRevD.78.074033} {\bibfield
  {journal} {\bibinfo  {journal} {Phys. Rev. D}\ }\textbf {\bibinfo {volume}
  {78}},\ \bibinfo {pages} {074033} (\bibinfo {year} {2008})},\ \Eprint
  {http://arxiv.org/abs/0808.3382} {arXiv:0808.3382 [hep-ph]} \BibitemShut
  {NoStop}%
\bibitem [{\citenamefont {Son}\ and\ \citenamefont
  {Zhitnitsky}(2004)}]{Son:2004tq}%
  \BibitemOpen
  \bibfield  {author} {\bibinfo {author} {\bibfnamefont {D.~T.}\ \bibnamefont
  {Son}}\ and\ \bibinfo {author} {\bibfnamefont {A.~R.}\ \bibnamefont
  {Zhitnitsky}},\ }\href {\doibase 10.1103/PhysRevD.70.074018} {\bibfield
  {journal} {\bibinfo  {journal} {Phys. Rev. D}\ }\textbf {\bibinfo {volume}
  {70}},\ \bibinfo {pages} {074018} (\bibinfo {year} {2004})},\ \Eprint
  {http://arxiv.org/abs/hep-ph/0405216} {arXiv:hep-ph/0405216} \BibitemShut
  {NoStop}%
\bibitem [{\citenamefont {Metlitski}\ and\ \citenamefont
  {Zhitnitsky}(2005)}]{Metlitski:2005pr}%
  \BibitemOpen
  \bibfield  {author} {\bibinfo {author} {\bibfnamefont {M.~A.}\ \bibnamefont
  {Metlitski}}\ and\ \bibinfo {author} {\bibfnamefont {A.~R.}\ \bibnamefont
  {Zhitnitsky}},\ }\href {\doibase 10.1103/PhysRevD.72.045011} {\bibfield
  {journal} {\bibinfo  {journal} {Phys. Rev. D}\ }\textbf {\bibinfo {volume}
  {72}},\ \bibinfo {pages} {045011} (\bibinfo {year} {2005})},\ \Eprint
  {http://arxiv.org/abs/hep-ph/0505072} {arXiv:hep-ph/0505072} \BibitemShut
  {NoStop}%
\bibitem [{\citenamefont {Kharzeev}\ \emph {et~al.}(2016)\citenamefont
  {Kharzeev}, \citenamefont {Liao}, \citenamefont {Voloshin},\ and\
  \citenamefont {Wang}}]{Kharzeev:2015znc}%
  \BibitemOpen
  \bibfield  {author} {\bibinfo {author} {\bibfnamefont {D.~E.}\ \bibnamefont
  {Kharzeev}}, \bibinfo {author} {\bibfnamefont {J.}~\bibnamefont {Liao}},
  \bibinfo {author} {\bibfnamefont {S.~A.}\ \bibnamefont {Voloshin}}, \ and\
  \bibinfo {author} {\bibfnamefont {G.}~\bibnamefont {Wang}},\ }\href {\doibase
  10.1016/j.ppnp.2016.01.001} {\bibfield  {journal} {\bibinfo  {journal} {Prog.
  Part. Nucl. Phys.}\ }\textbf {\bibinfo {volume} {88}},\ \bibinfo {pages} {1}
  (\bibinfo {year} {2016})},\ \Eprint {http://arxiv.org/abs/1511.04050}
  {arXiv:1511.04050 [hep-ph]} \BibitemShut {NoStop}%
\bibitem [{\citenamefont {Adare}\ \emph {et~al.}(2011)\citenamefont {Adare}
  \emph {et~al.}}]{PHENIX:2011rvu}%
  \BibitemOpen
  \bibfield  {author} {\bibinfo {author} {\bibfnamefont {A.}~\bibnamefont
  {Adare}} \emph {et~al.} (\bibinfo {collaboration} {PHENIX}),\ }\href
  {\doibase 10.1103/PhysRevC.83.064903} {\bibfield  {journal} {\bibinfo
  {journal} {Phys. Rev. C}\ }\textbf {\bibinfo {volume} {83}},\ \bibinfo
  {pages} {064903} (\bibinfo {year} {2011})},\ \Eprint
  {http://arxiv.org/abs/1102.0753} {arXiv:1102.0753 [nucl-ex]} \BibitemShut
  {NoStop}%
\bibitem [{\citenamefont {Aamodt}\ \emph {et~al.}(2011)\citenamefont {Aamodt}
  \emph {et~al.}}]{ALICE:2011gmo}%
  \BibitemOpen
  \bibfield  {author} {\bibinfo {author} {\bibfnamefont {K.}~\bibnamefont
  {Aamodt}} \emph {et~al.} (\bibinfo {collaboration} {ALICE}),\ }\href
  {\doibase 10.1140/epjc/s10052-011-1655-9} {\bibfield  {journal} {\bibinfo
  {journal} {Eur. Phys. J. C}\ }\textbf {\bibinfo {volume} {71}},\ \bibinfo
  {pages} {1655} (\bibinfo {year} {2011})},\ \Eprint
  {http://arxiv.org/abs/1101.4110} {arXiv:1101.4110 [hep-ex]} \BibitemShut
  {NoStop}%
\bibitem [{\citenamefont {Khachatryan}\ \emph {et~al.}(2011)\citenamefont
  {Khachatryan} \emph {et~al.}}]{CMS:2011jlm}%
  \BibitemOpen
  \bibfield  {author} {\bibinfo {author} {\bibfnamefont {V.}~\bibnamefont
  {Khachatryan}} \emph {et~al.} (\bibinfo {collaboration} {CMS}),\ }\href
  {\doibase 10.1007/JHEP05(2011)064} {\bibfield  {journal} {\bibinfo  {journal}
  {JHEP}\ }\textbf {\bibinfo {volume} {05}},\ \bibinfo {pages} {064} (\bibinfo
  {year} {2011})},\ \Eprint {http://arxiv.org/abs/1102.4282} {arXiv:1102.4282
  [hep-ex]} \BibitemShut {NoStop}%
\bibitem [{\citenamefont {Patra}\ \emph {et~al.}(2021)\citenamefont {Patra},
  \citenamefont {Mohanty},\ and\ \citenamefont {Nayak}}]{Patra:2020gzw}%
  \BibitemOpen
  \bibfield  {author} {\bibinfo {author} {\bibfnamefont {R.~N.}\ \bibnamefont
  {Patra}}, \bibinfo {author} {\bibfnamefont {B.}~\bibnamefont {Mohanty}}, \
  and\ \bibinfo {author} {\bibfnamefont {T.~K.}\ \bibnamefont {Nayak}},\ }\href
  {\doibase 10.1140/epjp/s13360-021-01660-0} {\bibfield  {journal} {\bibinfo
  {journal} {Eur. Phys. J. Plus}\ }\textbf {\bibinfo {volume} {136}},\ \bibinfo
  {pages} {702} (\bibinfo {year} {2021})},\ \Eprint
  {http://arxiv.org/abs/2008.02559} {arXiv:2008.02559 [hep-ph]} \BibitemShut
  {NoStop}%
\bibitem [{\citenamefont {Pradhan}\ \emph {et~al.}(2024)\citenamefont
  {Pradhan}, \citenamefont {Sahu}, \citenamefont {Rath}, \citenamefont
  {Sahoo},\ and\ \citenamefont {Cleymans}}]{Pradhan:2023bld}%
  \BibitemOpen
  \bibfield  {author} {\bibinfo {author} {\bibfnamefont {G.~S.}\ \bibnamefont
  {Pradhan}}, \bibinfo {author} {\bibfnamefont {D.}~\bibnamefont {Sahu}},
  \bibinfo {author} {\bibfnamefont {R.}~\bibnamefont {Rath}}, \bibinfo {author}
  {\bibfnamefont {R.}~\bibnamefont {Sahoo}}, \ and\ \bibinfo {author}
  {\bibfnamefont {J.}~\bibnamefont {Cleymans}},\ }\href {\doibase
  10.1140/epja/s10050-024-01270-1} {\bibfield  {journal} {\bibinfo  {journal}
  {Eur. Phys. J. A}\ }\textbf {\bibinfo {volume} {60}},\ \bibinfo {pages} {52}
  (\bibinfo {year} {2024})},\ \Eprint {http://arxiv.org/abs/2301.04038}
  {arXiv:2301.04038 [hep-ph]} \BibitemShut {NoStop}%
\bibitem [{\citenamefont {Tsallis}(1988)}]{Tsallis:1987eu}%
  \BibitemOpen
  \bibfield  {author} {\bibinfo {author} {\bibfnamefont {C.}~\bibnamefont
  {Tsallis}},\ }\href {\doibase 10.1007/BF01016429} {\bibfield  {journal}
  {\bibinfo  {journal} {J. Statist. Phys.}\ }\textbf {\bibinfo {volume} {52}},\
  \bibinfo {pages} {479} (\bibinfo {year} {1988})}\BibitemShut {NoStop}%
\bibitem [{\citenamefont {Tsallis}(2009)}]{Tsallis:2009zex}%
  \BibitemOpen
  \bibfield  {author} {\bibinfo {author} {\bibfnamefont {C.}~\bibnamefont
  {Tsallis}},\ }\href {\doibase 10.1007/978-0-387-85359-8} {\emph {\bibinfo
  {title} {{Introduction to Nonextensive Statistical Mechanics}: {Approaching a
  Complex World}}}}\ (\bibinfo  {publisher} {Springer},\ \bibinfo {address}
  {New York},\ \bibinfo {year} {2009})\BibitemShut {NoStop}%
\bibitem [{\citenamefont {Tsallis}\ \emph {et~al.}(1998)\citenamefont
  {Tsallis}, \citenamefont {Mendes},\ and\ \citenamefont
  {Plastino}}]{Tsallis:1998ws}%
  \BibitemOpen
  \bibfield  {author} {\bibinfo {author} {\bibfnamefont {C.}~\bibnamefont
  {Tsallis}}, \bibinfo {author} {\bibfnamefont {R.~S.}\ \bibnamefont {Mendes}},
  \ and\ \bibinfo {author} {\bibfnamefont {A.~R.}\ \bibnamefont {Plastino}},\
  }\href {\doibase 10.1016/S0378-4371(98)00437-3} {\bibfield  {journal}
  {\bibinfo  {journal} {Physica A}\ }\textbf {\bibinfo {volume} {261}},\
  \bibinfo {pages} {534} (\bibinfo {year} {1998})}\BibitemShut {NoStop}%
\bibitem [{\citenamefont {Biro}\ \emph {et~al.}(2009)\citenamefont {Biro},
  \citenamefont {Purcsel},\ and\ \citenamefont {Urmossy}}]{Biro:2008hz}%
  \BibitemOpen
  \bibfield  {author} {\bibinfo {author} {\bibfnamefont {T.~S.}\ \bibnamefont
  {Biro}}, \bibinfo {author} {\bibfnamefont {G.}~\bibnamefont {Purcsel}}, \
  and\ \bibinfo {author} {\bibfnamefont {K.}~\bibnamefont {Urmossy}},\ }\href
  {\doibase 10.1140/epja/i2009-10806-6} {\bibfield  {journal} {\bibinfo
  {journal} {Eur. Phys. J. A}\ }\textbf {\bibinfo {volume} {40}},\ \bibinfo
  {pages} {325} (\bibinfo {year} {2009})},\ \Eprint
  {http://arxiv.org/abs/0812.2104} {arXiv:0812.2104 [hep-ph]} \BibitemShut
  {NoStop}%
\bibitem [{\citenamefont {Wilk}\ and\ \citenamefont
  {Wlodarczyk}(2009)}]{Wilk:2008ue}%
  \BibitemOpen
  \bibfield  {author} {\bibinfo {author} {\bibfnamefont {G.}~\bibnamefont
  {Wilk}}\ and\ \bibinfo {author} {\bibfnamefont {Z.}~\bibnamefont
  {Wlodarczyk}},\ }\href {\doibase 10.1140/epja/i2009-10803-9} {\bibfield
  {journal} {\bibinfo  {journal} {Eur. Phys. J. A}\ }\textbf {\bibinfo {volume}
  {40}},\ \bibinfo {pages} {299} (\bibinfo {year} {2009})},\ \Eprint
  {http://arxiv.org/abs/0810.2939} {arXiv:0810.2939 [hep-ph]} \BibitemShut
  {NoStop}%
\bibitem [{\citenamefont {Pereira}\ \emph {et~al.}(2009)\citenamefont
  {Pereira}, \citenamefont {Silva},\ and\ \citenamefont
  {Alcaniz}}]{Pereira:2009ja}%
  \BibitemOpen
  \bibfield  {author} {\bibinfo {author} {\bibfnamefont {F.~I.~M.}\
  \bibnamefont {Pereira}}, \bibinfo {author} {\bibfnamefont {R.}~\bibnamefont
  {Silva}}, \ and\ \bibinfo {author} {\bibfnamefont {J.~S.}\ \bibnamefont
  {Alcaniz}},\ }\href {\doibase 10.1016/j.physleta.2009.09.046} {\bibfield
  {journal} {\bibinfo  {journal} {Phys. Lett. A}\ }\textbf {\bibinfo {volume}
  {373}},\ \bibinfo {pages} {4214} (\bibinfo {year} {2009})},\ \Eprint
  {http://arxiv.org/abs/0906.2422} {arXiv:0906.2422 [nucl-th]} \BibitemShut
  {NoStop}%
\bibitem [{\citenamefont {Conroy}\ \emph {et~al.}(2010)\citenamefont {Conroy},
  \citenamefont {Miller},\ and\ \citenamefont {Plastino}}]{Conroy:2010wt}%
  \BibitemOpen
  \bibfield  {author} {\bibinfo {author} {\bibfnamefont {J.~M.}\ \bibnamefont
  {Conroy}}, \bibinfo {author} {\bibfnamefont {H.~G.}\ \bibnamefont {Miller}},
  \ and\ \bibinfo {author} {\bibfnamefont {A.~R.}\ \bibnamefont {Plastino}},\
  }\href {\doibase 10.1016/j.physleta.2010.09.038} {\bibfield  {journal}
  {\bibinfo  {journal} {Phys. Lett. A}\ }\textbf {\bibinfo {volume} {374}},\
  \bibinfo {pages} {4581} (\bibinfo {year} {2010})},\ \Eprint
  {http://arxiv.org/abs/1006.3963} {arXiv:1006.3963 [cond-mat.stat-mech]}
  \BibitemShut {NoStop}%
\bibitem [{\citenamefont {Cleymans}\ and\ \citenamefont
  {Worku}(2012{\natexlab{a}})}]{Cleymans:2011in}%
  \BibitemOpen
  \bibfield  {author} {\bibinfo {author} {\bibfnamefont {J.}~\bibnamefont
  {Cleymans}}\ and\ \bibinfo {author} {\bibfnamefont {D.}~\bibnamefont
  {Worku}},\ }\href {\doibase 10.1088/0954-3899/39/2/025006} {\bibfield
  {journal} {\bibinfo  {journal} {J. Phys. G}\ }\textbf {\bibinfo {volume}
  {39}},\ \bibinfo {pages} {025006} (\bibinfo {year} {2012}{\natexlab{a}})},\
  \Eprint {http://arxiv.org/abs/1110.5526} {arXiv:1110.5526 [hep-ph]}
  \BibitemShut {NoStop}%
\bibitem [{\citenamefont {Cleymans}\ and\ \citenamefont
  {Worku}(2012{\natexlab{b}})}]{Cleymans:2012ya}%
  \BibitemOpen
  \bibfield  {author} {\bibinfo {author} {\bibfnamefont {J.}~\bibnamefont
  {Cleymans}}\ and\ \bibinfo {author} {\bibfnamefont {D.}~\bibnamefont
  {Worku}},\ }\href {\doibase 10.1140/epja/i2012-12160-0} {\bibfield  {journal}
  {\bibinfo  {journal} {Eur. Phys. J. A}\ }\textbf {\bibinfo {volume} {48}},\
  \bibinfo {pages} {160} (\bibinfo {year} {2012}{\natexlab{b}})},\ \Eprint
  {http://arxiv.org/abs/1203.4343} {arXiv:1203.4343 [hep-ph]} \BibitemShut
  {NoStop}%
\bibitem [{\citenamefont {Kapusta}(2021)}]{Kapusta:2021zfo}%
  \BibitemOpen
  \bibfield  {author} {\bibinfo {author} {\bibfnamefont {J.~I.}\ \bibnamefont
  {Kapusta}},\ }\href {\doibase 10.1142/S021830132130006X} {\bibfield
  {journal} {\bibinfo  {journal} {Int. J. Mod. Phys. E}\ }\textbf {\bibinfo
  {volume} {30}},\ \bibinfo {pages} {2130006} (\bibinfo {year} {2021})},\
  \Eprint {http://arxiv.org/abs/2106.06824} {arXiv:2106.06824 [hep-ph]}
  \BibitemShut {NoStop}%
\bibitem [{\citenamefont {Pereira}\ \emph {et~al.}(2007)\citenamefont
  {Pereira}, \citenamefont {SIlva},\ and\ \citenamefont
  {Alcaniz}}]{Pereira:2007hp}%
  \BibitemOpen
  \bibfield  {author} {\bibinfo {author} {\bibfnamefont {F.~I.~M.}\
  \bibnamefont {Pereira}}, \bibinfo {author} {\bibfnamefont {R.}~\bibnamefont
  {SIlva}}, \ and\ \bibinfo {author} {\bibfnamefont {J.~S.}\ \bibnamefont
  {Alcaniz}},\ }\href {\doibase 10.1103/PhysRevC.76.015201} {\bibfield
  {journal} {\bibinfo  {journal} {Phys. Rev. C}\ }\textbf {\bibinfo {volume}
  {76}},\ \bibinfo {pages} {015201} (\bibinfo {year} {2007})},\ \Eprint
  {http://arxiv.org/abs/0705.0300} {arXiv:0705.0300 [nucl-th]} \BibitemShut
  {NoStop}%
\bibitem [{\citenamefont {Rozynek}\ and\ \citenamefont
  {Wilk}(2009)}]{Rozynek:2009zh}%
  \BibitemOpen
  \bibfield  {author} {\bibinfo {author} {\bibfnamefont {J.}~\bibnamefont
  {Rozynek}}\ and\ \bibinfo {author} {\bibfnamefont {G.}~\bibnamefont {Wilk}},\
  }\href {\doibase 10.1088/0954-3899/36/12/125108} {\bibfield  {journal}
  {\bibinfo  {journal} {J. Phys. G}\ }\textbf {\bibinfo {volume} {36}},\
  \bibinfo {pages} {125108} (\bibinfo {year} {2009})},\ \Eprint
  {http://arxiv.org/abs/0905.3408} {arXiv:0905.3408 [nucl-th]} \BibitemShut
  {NoStop}%
\bibitem [{\citenamefont {Ro\.zynek}\ and\ \citenamefont
  {Wilk}(2016)}]{Rozynek:2015zca}%
  \BibitemOpen
  \bibfield  {author} {\bibinfo {author} {\bibfnamefont {J.}~\bibnamefont
  {Ro\.zynek}}\ and\ \bibinfo {author} {\bibfnamefont {G.}~\bibnamefont
  {Wilk}},\ }\href {\doibase 10.1140/epja/i2016-16013-6} {\bibfield  {journal}
  {\bibinfo  {journal} {Eur. Phys. J. A}\ }\textbf {\bibinfo {volume} {52}},\
  \bibinfo {pages} {13} (\bibinfo {year} {2016})},\ \bibinfo {note} {[Erratum:
  Eur.Phys.J.A 52, 204 (2016)]},\ \Eprint {http://arxiv.org/abs/1510.08516}
  {arXiv:1510.08516 [nucl-th]} \BibitemShut {NoStop}%
\bibitem [{\citenamefont {Zhao}(2020)}]{Zhao:2020xob}%
  \BibitemOpen
  \bibfield  {author} {\bibinfo {author} {\bibfnamefont {Y.-P.}\ \bibnamefont
  {Zhao}},\ }\href {\doibase 10.1103/PhysRevD.101.096006} {\bibfield  {journal}
  {\bibinfo  {journal} {Phys. Rev. D}\ }\textbf {\bibinfo {volume} {101}},\
  \bibinfo {pages} {096006} (\bibinfo {year} {2020})},\ \Eprint
  {http://arxiv.org/abs/2004.14556} {arXiv:2004.14556 [hep-ph]} \BibitemShut
  {NoStop}%
\bibitem [{\citenamefont {Pradhan}\ \emph {et~al.}(2023)\citenamefont
  {Pradhan}, \citenamefont {Sahu}, \citenamefont {Deb},\ and\ \citenamefont
  {Sahoo}}]{Pradhan:2021vtp}%
  \BibitemOpen
  \bibfield  {author} {\bibinfo {author} {\bibfnamefont {G.~S.}\ \bibnamefont
  {Pradhan}}, \bibinfo {author} {\bibfnamefont {D.}~\bibnamefont {Sahu}},
  \bibinfo {author} {\bibfnamefont {S.}~\bibnamefont {Deb}}, \ and\ \bibinfo
  {author} {\bibfnamefont {R.}~\bibnamefont {Sahoo}},\ }\href {\doibase
  10.1088/1361-6471/acc478} {\bibfield  {journal} {\bibinfo  {journal} {J.
  Phys. G}\ }\textbf {\bibinfo {volume} {50}},\ \bibinfo {pages} {055104}
  (\bibinfo {year} {2023})},\ \Eprint {http://arxiv.org/abs/2106.14297}
  {arXiv:2106.14297 [hep-ph]} \BibitemShut {NoStop}%
\bibitem [{\citenamefont {Zhao}\ \emph {et~al.}(2023)\citenamefont {Zhao},
  \citenamefont {Wang}, \citenamefont {Zuo},\ and\ \citenamefont
  {Li}}]{Zhao:2023xpj}%
  \BibitemOpen
  \bibfield  {author} {\bibinfo {author} {\bibfnamefont {Y.-P.}\ \bibnamefont
  {Zhao}}, \bibinfo {author} {\bibfnamefont {C.-Y.}\ \bibnamefont {Wang}},
  \bibinfo {author} {\bibfnamefont {S.-Y.}\ \bibnamefont {Zuo}}, \ and\
  \bibinfo {author} {\bibfnamefont {C.-M.}\ \bibnamefont {Li}},\ }\href
  {\doibase 10.1088/1674-1137/acbf2a} {\bibfield  {journal} {\bibinfo
  {journal} {Chin. Phys. C}\ }\textbf {\bibinfo {volume} {47}},\ \bibinfo
  {pages} {053103} (\bibinfo {year} {2023})},\ \Eprint
  {http://arxiv.org/abs/2302.12010} {arXiv:2302.12010 [hep-ph]} \BibitemShut
  {NoStop}%
\bibitem [{\citenamefont {Farias}\ \emph {et~al.}(2014)\citenamefont {Farias},
  \citenamefont {Gomes}, \citenamefont {Krein},\ and\ \citenamefont
  {Pinto}}]{Farias:2014eca}%
  \BibitemOpen
  \bibfield  {author} {\bibinfo {author} {\bibfnamefont {R.~L.~S.}\
  \bibnamefont {Farias}}, \bibinfo {author} {\bibfnamefont {K.~P.}\
  \bibnamefont {Gomes}}, \bibinfo {author} {\bibfnamefont {G.~I.}\ \bibnamefont
  {Krein}}, \ and\ \bibinfo {author} {\bibfnamefont {M.~B.}\ \bibnamefont
  {Pinto}},\ }\href {\doibase 10.1103/PhysRevC.90.025203} {\bibfield  {journal}
  {\bibinfo  {journal} {Phys. Rev. C}\ }\textbf {\bibinfo {volume} {90}},\
  \bibinfo {pages} {025203} (\bibinfo {year} {2014})},\ \Eprint
  {http://arxiv.org/abs/1404.3931} {arXiv:1404.3931 [hep-ph]} \BibitemShut
  {NoStop}%
\bibitem [{\citenamefont {Ferreira}\ \emph {et~al.}(2014)\citenamefont
  {Ferreira}, \citenamefont {Costa}, \citenamefont {Louren\c{c}o},
  \citenamefont {Frederico},\ and\ \citenamefont
  {Provid\^encia}}]{Ferreira:2014kpa}%
  \BibitemOpen
  \bibfield  {author} {\bibinfo {author} {\bibfnamefont {M.}~\bibnamefont
  {Ferreira}}, \bibinfo {author} {\bibfnamefont {P.}~\bibnamefont {Costa}},
  \bibinfo {author} {\bibfnamefont {O.}~\bibnamefont {Louren\c{c}o}}, \bibinfo
  {author} {\bibfnamefont {T.}~\bibnamefont {Frederico}}, \ and\ \bibinfo
  {author} {\bibfnamefont {C.}~\bibnamefont {Provid\^encia}},\ }\href {\doibase
  10.1103/PhysRevD.89.116011} {\bibfield  {journal} {\bibinfo  {journal} {Phys.
  Rev. D}\ }\textbf {\bibinfo {volume} {89}},\ \bibinfo {pages} {116011}
  (\bibinfo {year} {2014})},\ \Eprint {http://arxiv.org/abs/1404.5577}
  {arXiv:1404.5577 [hep-ph]} \BibitemShut {NoStop}%
\bibitem [{\citenamefont {Ayala}\ \emph {et~al.}(2015)\citenamefont {Ayala},
  \citenamefont {Loewe},\ and\ \citenamefont {Zamora}}]{Ayala:2014gwa}%
  \BibitemOpen
  \bibfield  {author} {\bibinfo {author} {\bibfnamefont {A.}~\bibnamefont
  {Ayala}}, \bibinfo {author} {\bibfnamefont {M.}~\bibnamefont {Loewe}}, \ and\
  \bibinfo {author} {\bibfnamefont {R.}~\bibnamefont {Zamora}},\ }\href
  {\doibase 10.1103/PhysRevD.91.016002} {\bibfield  {journal} {\bibinfo
  {journal} {Phys. Rev. D}\ }\textbf {\bibinfo {volume} {91}},\ \bibinfo
  {pages} {016002} (\bibinfo {year} {2015})},\ \Eprint
  {http://arxiv.org/abs/1406.7408} {arXiv:1406.7408 [hep-ph]} \BibitemShut
  {NoStop}%
\bibitem [{\citenamefont {Farias}\ \emph {et~al.}(2017)\citenamefont {Farias},
  \citenamefont {Timoteo}, \citenamefont {Avancini}, \citenamefont {Pinto},\
  and\ \citenamefont {Krein}}]{Farias:2016gmy}%
  \BibitemOpen
  \bibfield  {author} {\bibinfo {author} {\bibfnamefont {R.~L.~S.}\
  \bibnamefont {Farias}}, \bibinfo {author} {\bibfnamefont {V.~S.}\
  \bibnamefont {Timoteo}}, \bibinfo {author} {\bibfnamefont {S.~S.}\
  \bibnamefont {Avancini}}, \bibinfo {author} {\bibfnamefont {M.~B.}\
  \bibnamefont {Pinto}}, \ and\ \bibinfo {author} {\bibfnamefont
  {G.}~\bibnamefont {Krein}},\ }\href {\doibase 10.1140/epja/i2017-12320-8}
  {\bibfield  {journal} {\bibinfo  {journal} {Eur. Phys. J. A}\ }\textbf
  {\bibinfo {volume} {53}},\ \bibinfo {pages} {101} (\bibinfo {year} {2017})},\
  \Eprint {http://arxiv.org/abs/1603.03847} {arXiv:1603.03847 [hep-ph]}
  \BibitemShut {NoStop}%
\bibitem [{\citenamefont {Tawfik}\ \emph {et~al.}(2016)\citenamefont {Tawfik},
  \citenamefont {Diab}, \citenamefont {Ezzelarab},\ and\ \citenamefont
  {Shalaby}}]{Tawfik:2016lih}%
  \BibitemOpen
  \bibfield  {author} {\bibinfo {author} {\bibfnamefont {A.~N.}\ \bibnamefont
  {Tawfik}}, \bibinfo {author} {\bibfnamefont {A.~M.}\ \bibnamefont {Diab}},
  \bibinfo {author} {\bibfnamefont {N.}~\bibnamefont {Ezzelarab}}, \ and\
  \bibinfo {author} {\bibfnamefont {A.~G.}\ \bibnamefont {Shalaby}},\ }\href
  {\doibase 10.1155/2016/1381479} {\bibfield  {journal} {\bibinfo  {journal}
  {Adv. High Energy Phys.}\ }\textbf {\bibinfo {volume} {2016}},\ \bibinfo
  {pages} {1381479} (\bibinfo {year} {2016})},\ \Eprint
  {http://arxiv.org/abs/1604.00043} {arXiv:1604.00043 [hep-ph]} \BibitemShut
  {NoStop}%
\bibitem [{\citenamefont {Tawfik}\ \emph
  {et~al.}(2018{\natexlab{a}})\citenamefont {Tawfik}, \citenamefont {Diab},\
  and\ \citenamefont {Hussein}}]{Tawfik:2016gye}%
  \BibitemOpen
  \bibfield  {author} {\bibinfo {author} {\bibfnamefont {A.~N.}\ \bibnamefont
  {Tawfik}}, \bibinfo {author} {\bibfnamefont {A.~M.}\ \bibnamefont {Diab}}, \
  and\ \bibinfo {author} {\bibfnamefont {M.~T.}\ \bibnamefont {Hussein}},\
  }\href {\doibase 10.1088/1361-6471/aaba9e} {\bibfield  {journal} {\bibinfo
  {journal} {J. Phys. G}\ }\textbf {\bibinfo {volume} {45}},\ \bibinfo {pages}
  {055008} (\bibinfo {year} {2018}{\natexlab{a}})},\ \Eprint
  {http://arxiv.org/abs/1604.08174} {arXiv:1604.08174 [hep-lat]} \BibitemShut
  {NoStop}%
\bibitem [{\citenamefont {Tawfik}\ \emph
  {et~al.}(2018{\natexlab{b}})\citenamefont {Tawfik}, \citenamefont {Diab},\
  and\ \citenamefont {Hussein}}]{Tawfik:2017cdx}%
  \BibitemOpen
  \bibfield  {author} {\bibinfo {author} {\bibfnamefont {A.~N.}\ \bibnamefont
  {Tawfik}}, \bibinfo {author} {\bibfnamefont {A.~M.}\ \bibnamefont {Diab}}, \
  and\ \bibinfo {author} {\bibfnamefont {M.~T.}\ \bibnamefont {Hussein}},\
  }\href {\doibase 10.1134/S1063776118050138} {\bibfield  {journal} {\bibinfo
  {journal} {J. Exp. Theor. Phys.}\ }\textbf {\bibinfo {volume} {126}},\
  \bibinfo {pages} {620} (\bibinfo {year} {2018}{\natexlab{b}})},\ \Eprint
  {http://arxiv.org/abs/1712.03264} {arXiv:1712.03264 [hep-ph]} \BibitemShut
  {NoStop}%
\bibitem [{\citenamefont {Tawfik}\ and\ \citenamefont
  {Diab}(2021)}]{Tawfik:2021eeb}%
  \BibitemOpen
  \bibfield  {author} {\bibinfo {author} {\bibfnamefont {A.~N.}\ \bibnamefont
  {Tawfik}}\ and\ \bibinfo {author} {\bibfnamefont {A.~M.}\ \bibnamefont
  {Diab}},\ }\href {\doibase 10.1140/epja/s10050-021-00501-z} {\bibfield
  {journal} {\bibinfo  {journal} {Eur. Phys. J. A}\ }\textbf {\bibinfo {volume}
  {57}},\ \bibinfo {pages} {200} (\bibinfo {year} {2021})},\ \Eprint
  {http://arxiv.org/abs/2106.04576} {arXiv:2106.04576 [hep-ph]} \BibitemShut
  {NoStop}%
\bibitem [{\citenamefont {Miransky}\ and\ \citenamefont
  {Shovkovy}(2002)}]{Miransky:2002rp}%
  \BibitemOpen
  \bibfield  {author} {\bibinfo {author} {\bibfnamefont {V.}~\bibnamefont
  {Miransky}}\ and\ \bibinfo {author} {\bibfnamefont {I.}~\bibnamefont
  {Shovkovy}},\ }\href {\doibase 10.1103/PhysRevD.66.045006} {\bibfield
  {journal} {\bibinfo  {journal} {Phys. Rev. D}\ }\textbf {\bibinfo {volume}
  {66}},\ \bibinfo {pages} {045006} (\bibinfo {year} {2002})},\ \Eprint
  {http://arxiv.org/abs/hep-ph/0205348} {arXiv:hep-ph/0205348} \BibitemShut
  {NoStop}%
\bibitem [{\citenamefont {Menezes}\ \emph {et~al.}(2009)\citenamefont
  {Menezes}, \citenamefont {Benghi~Pinto}, \citenamefont {Avancini},
  \citenamefont {Perez~Martinez},\ and\ \citenamefont
  {Providencia}}]{Menezes:2008qt}%
  \BibitemOpen
  \bibfield  {author} {\bibinfo {author} {\bibfnamefont {D.~P.}\ \bibnamefont
  {Menezes}}, \bibinfo {author} {\bibfnamefont {M.}~\bibnamefont
  {Benghi~Pinto}}, \bibinfo {author} {\bibfnamefont {S.~S.}\ \bibnamefont
  {Avancini}}, \bibinfo {author} {\bibfnamefont {A.}~\bibnamefont
  {Perez~Martinez}}, \ and\ \bibinfo {author} {\bibfnamefont {C.}~\bibnamefont
  {Providencia}},\ }\href {\doibase 10.1103/PhysRevC.79.035807} {\bibfield
  {journal} {\bibinfo  {journal} {Phys. Rev. C}\ }\textbf {\bibinfo {volume}
  {79}},\ \bibinfo {pages} {035807} (\bibinfo {year} {2009})},\ \Eprint
  {http://arxiv.org/abs/0811.3361} {arXiv:0811.3361 [nucl-th]} \BibitemShut
  {NoStop}%
\bibitem [{\citenamefont {Boomsma}\ and\ \citenamefont
  {Boer}(2010)}]{Boomsma:2009yk}%
  \BibitemOpen
  \bibfield  {author} {\bibinfo {author} {\bibfnamefont {J.~K.}\ \bibnamefont
  {Boomsma}}\ and\ \bibinfo {author} {\bibfnamefont {D.}~\bibnamefont {Boer}},\
  }\href {\doibase 10.1103/PhysRevD.81.074005} {\bibfield  {journal} {\bibinfo
  {journal} {Phys. Rev. D}\ }\textbf {\bibinfo {volume} {81}},\ \bibinfo
  {pages} {074005} (\bibinfo {year} {2010})},\ \Eprint
  {http://arxiv.org/abs/0911.2164} {arXiv:0911.2164 [hep-ph]} \BibitemShut
  {NoStop}%
\bibitem [{\citenamefont {Chatterjee}\ \emph {et~al.}(2011)\citenamefont
  {Chatterjee}, \citenamefont {Mishra},\ and\ \citenamefont
  {Mishra}}]{Chatterjee:2011ry}%
  \BibitemOpen
  \bibfield  {author} {\bibinfo {author} {\bibfnamefont {B.}~\bibnamefont
  {Chatterjee}}, \bibinfo {author} {\bibfnamefont {H.}~\bibnamefont {Mishra}},
  \ and\ \bibinfo {author} {\bibfnamefont {A.}~\bibnamefont {Mishra}},\ }\href
  {\doibase 10.1103/PhysRevD.84.014016} {\bibfield  {journal} {\bibinfo
  {journal} {Phys. Rev. D}\ }\textbf {\bibinfo {volume} {84}},\ \bibinfo
  {pages} {014016} (\bibinfo {year} {2011})},\ \Eprint
  {http://arxiv.org/abs/1101.0498} {arXiv:1101.0498 [hep-ph]} \BibitemShut
  {NoStop}%
\bibitem [{\citenamefont {Avancini}\ \emph {et~al.}(2011)\citenamefont
  {Avancini}, \citenamefont {Menezes},\ and\ \citenamefont
  {Providencia}}]{Avancini:2011zz}%
  \BibitemOpen
  \bibfield  {author} {\bibinfo {author} {\bibfnamefont {S.~S.}\ \bibnamefont
  {Avancini}}, \bibinfo {author} {\bibfnamefont {D.~P.}\ \bibnamefont
  {Menezes}}, \ and\ \bibinfo {author} {\bibfnamefont {C.}~\bibnamefont
  {Providencia}},\ }\href {\doibase 10.1103/PhysRevC.83.065805} {\bibfield
  {journal} {\bibinfo  {journal} {Phys. Rev. C}\ }\textbf {\bibinfo {volume}
  {83}},\ \bibinfo {pages} {065805} (\bibinfo {year} {2011})}\BibitemShut
  {NoStop}%
\bibitem [{\citenamefont {Ferrer}\ \emph {et~al.}(2015)\citenamefont {Ferrer},
  \citenamefont {de~la Incera},\ and\ \citenamefont {Wen}}]{Ferrer:2014qka}%
  \BibitemOpen
  \bibfield  {author} {\bibinfo {author} {\bibfnamefont {E.}~\bibnamefont
  {Ferrer}}, \bibinfo {author} {\bibfnamefont {V.}~\bibnamefont {de~la
  Incera}}, \ and\ \bibinfo {author} {\bibfnamefont {X.}~\bibnamefont {Wen}},\
  }\href {\doibase 10.1103/PhysRevD.91.054006} {\bibfield  {journal} {\bibinfo
  {journal} {Phys. Rev. D}\ }\textbf {\bibinfo {volume} {91}},\ \bibinfo
  {pages} {054006} (\bibinfo {year} {2015})},\ \Eprint
  {http://arxiv.org/abs/1407.3503} {arXiv:1407.3503 [nucl-th]} \BibitemShut
  {NoStop}%
\bibitem [{\citenamefont {Yu}\ \emph {et~al.}(2015)\citenamefont {Yu},
  \citenamefont {Van~Doorsselaere},\ and\ \citenamefont {Huang}}]{Yu:2014xoa}%
  \BibitemOpen
  \bibfield  {author} {\bibinfo {author} {\bibfnamefont {L.}~\bibnamefont
  {Yu}}, \bibinfo {author} {\bibfnamefont {J.}~\bibnamefont
  {Van~Doorsselaere}}, \ and\ \bibinfo {author} {\bibfnamefont
  {M.}~\bibnamefont {Huang}},\ }\href {\doibase 10.1103/PhysRevD.91.074011}
  {\bibfield  {journal} {\bibinfo  {journal} {Phys. Rev. D}\ }\textbf {\bibinfo
  {volume} {91}},\ \bibinfo {pages} {074011} (\bibinfo {year} {2015})},\
  \Eprint {http://arxiv.org/abs/1411.7552} {arXiv:1411.7552 [hep-ph]}
  \BibitemShut {NoStop}%
\bibitem [{\citenamefont {Mao}(2016)}]{Mao:2016fha}%
  \BibitemOpen
  \bibfield  {author} {\bibinfo {author} {\bibfnamefont {S.}~\bibnamefont
  {Mao}},\ }\href {\doibase 10.1016/j.physletb.2016.05.018} {\bibfield
  {journal} {\bibinfo  {journal} {Phys. Lett. B}\ }\textbf {\bibinfo {volume}
  {758}},\ \bibinfo {pages} {195} (\bibinfo {year} {2016})},\ \Eprint
  {http://arxiv.org/abs/1602.06503} {arXiv:1602.06503 [hep-ph]} \BibitemShut
  {NoStop}%
\bibitem [{\citenamefont {Klevansky}(1992)}]{Klevansky:1992qe}%
  \BibitemOpen
  \bibfield  {author} {\bibinfo {author} {\bibfnamefont {S.~P.}\ \bibnamefont
  {Klevansky}},\ }\href {\doibase 10.1103/RevModPhys.64.649} {\bibfield
  {journal} {\bibinfo  {journal} {Rev. Mod. Phys.}\ }\textbf {\bibinfo {volume}
  {64}},\ \bibinfo {pages} {649} (\bibinfo {year} {1992})}\BibitemShut
  {NoStop}%
\bibitem [{\citenamefont {Hatsuda}\ and\ \citenamefont
  {Kunihiro}(1994)}]{Hatsuda:1994pi}%
  \BibitemOpen
  \bibfield  {author} {\bibinfo {author} {\bibfnamefont {T.}~\bibnamefont
  {Hatsuda}}\ and\ \bibinfo {author} {\bibfnamefont {T.}~\bibnamefont
  {Kunihiro}},\ }\href {\doibase 10.1016/0370-1573(94)90022-1} {\bibfield
  {journal} {\bibinfo  {journal} {Phys. Rept.}\ }\textbf {\bibinfo {volume}
  {247}},\ \bibinfo {pages} {221} (\bibinfo {year} {1994})},\ \Eprint
  {http://arxiv.org/abs/hep-ph/9401310} {arXiv:hep-ph/9401310} \BibitemShut
  {NoStop}%
\bibitem [{\citenamefont {Buballa}(2005)}]{Buballa:2003qv}%
  \BibitemOpen
  \bibfield  {author} {\bibinfo {author} {\bibfnamefont {M.}~\bibnamefont
  {Buballa}},\ }\href {\doibase 10.1016/j.physrep.2004.11.004} {\bibfield
  {journal} {\bibinfo  {journal} {Phys. Rept.}\ }\textbf {\bibinfo {volume}
  {407}},\ \bibinfo {pages} {205} (\bibinfo {year} {2005})},\ \Eprint
  {http://arxiv.org/abs/hep-ph/0402234} {arXiv:hep-ph/0402234} \BibitemShut
  {NoStop}%
\bibitem [{\citenamefont {Rehberg}\ \emph {et~al.}(1996)\citenamefont
  {Rehberg}, \citenamefont {Klevansky},\ and\ \citenamefont
  {Hufner}}]{Rehberg:1995kh}%
  \BibitemOpen
  \bibfield  {author} {\bibinfo {author} {\bibfnamefont {P.}~\bibnamefont
  {Rehberg}}, \bibinfo {author} {\bibfnamefont {S.~P.}\ \bibnamefont
  {Klevansky}}, \ and\ \bibinfo {author} {\bibfnamefont {J.}~\bibnamefont
  {Hufner}},\ }\href {\doibase 10.1103/PhysRevC.53.410} {\bibfield  {journal}
  {\bibinfo  {journal} {Phys. Rev. C}\ }\textbf {\bibinfo {volume} {53}},\
  \bibinfo {pages} {410} (\bibinfo {year} {1996})},\ \Eprint
  {http://arxiv.org/abs/hep-ph/9506436} {arXiv:hep-ph/9506436} \BibitemShut
  {NoStop}%
\bibitem [{\citenamefont {Biro}\ and\ \citenamefont
  {Purcsel}(2005)}]{Biro:2005uv}%
  \BibitemOpen
  \bibfield  {author} {\bibinfo {author} {\bibfnamefont {T.~S.}\ \bibnamefont
  {Biro}}\ and\ \bibinfo {author} {\bibfnamefont {G.}~\bibnamefont {Purcsel}},\
  }\href {\doibase 10.1103/PhysRevLett.95.162302} {\bibfield  {journal}
  {\bibinfo  {journal} {Phys. Rev. Lett.}\ }\textbf {\bibinfo {volume} {95}},\
  \bibinfo {pages} {162302} (\bibinfo {year} {2005})},\ \Eprint
  {http://arxiv.org/abs/hep-ph/0503204} {arXiv:hep-ph/0503204} \BibitemShut
  {NoStop}%
\bibitem [{\citenamefont {Gusynin}\ \emph {et~al.}(1996)\citenamefont
  {Gusynin}, \citenamefont {Miransky},\ and\ \citenamefont
  {Shovkovy}}]{Gusynin:1995nb}%
  \BibitemOpen
  \bibfield  {author} {\bibinfo {author} {\bibfnamefont {V.~P.}\ \bibnamefont
  {Gusynin}}, \bibinfo {author} {\bibfnamefont {V.~A.}\ \bibnamefont
  {Miransky}}, \ and\ \bibinfo {author} {\bibfnamefont {I.~A.}\ \bibnamefont
  {Shovkovy}},\ }\href {\doibase 10.1016/0550-3213(96)00021-1} {\bibfield
  {journal} {\bibinfo  {journal} {Nucl. Phys. B}\ }\textbf {\bibinfo {volume}
  {462}},\ \bibinfo {pages} {249} (\bibinfo {year} {1996})},\ \Eprint
  {http://arxiv.org/abs/hep-ph/9509320} {arXiv:hep-ph/9509320} \BibitemShut
  {NoStop}%
\bibitem [{\citenamefont {Gusynin}\ and\ \citenamefont
  {Shovkovy}(1997)}]{Gusynin:1997kj}%
  \BibitemOpen
  \bibfield  {author} {\bibinfo {author} {\bibfnamefont {V.~P.}\ \bibnamefont
  {Gusynin}}\ and\ \bibinfo {author} {\bibfnamefont {I.~A.}\ \bibnamefont
  {Shovkovy}},\ }\href {\doibase 10.1103/PhysRevD.56.5251} {\bibfield
  {journal} {\bibinfo  {journal} {Phys. Rev. D}\ }\textbf {\bibinfo {volume}
  {56}},\ \bibinfo {pages} {5251} (\bibinfo {year} {1997})},\ \Eprint
  {http://arxiv.org/abs/hep-ph/9704394} {arXiv:hep-ph/9704394} \BibitemShut
  {NoStop}%
\bibitem [{\citenamefont {D'Elia}\ \emph {et~al.}(2018)\citenamefont {D'Elia},
  \citenamefont {Manigrasso}, \citenamefont {Negro},\ and\ \citenamefont
  {Sanfilippo}}]{DElia:2018xwo}%
  \BibitemOpen
  \bibfield  {author} {\bibinfo {author} {\bibfnamefont {M.}~\bibnamefont
  {D'Elia}}, \bibinfo {author} {\bibfnamefont {F.}~\bibnamefont {Manigrasso}},
  \bibinfo {author} {\bibfnamefont {F.}~\bibnamefont {Negro}}, \ and\ \bibinfo
  {author} {\bibfnamefont {F.}~\bibnamefont {Sanfilippo}},\ }\href {\doibase
  10.1103/PhysRevD.98.054509} {\bibfield  {journal} {\bibinfo  {journal} {Phys.
  Rev. D}\ }\textbf {\bibinfo {volume} {98}},\ \bibinfo {pages} {054509}
  (\bibinfo {year} {2018})},\ \Eprint {http://arxiv.org/abs/1808.07008}
  {arXiv:1808.07008 [hep-lat]} \BibitemShut {NoStop}%
\bibitem [{\citenamefont {Endrodi}(2015)}]{Endrodi:2015oba}%
  \BibitemOpen
  \bibfield  {author} {\bibinfo {author} {\bibfnamefont {G.}~\bibnamefont
  {Endrodi}},\ }\href {\doibase 10.1007/JHEP07(2015)173} {\bibfield  {journal}
  {\bibinfo  {journal} {JHEP}\ }\textbf {\bibinfo {volume} {07}},\ \bibinfo
  {pages} {173} (\bibinfo {year} {2015})},\ \Eprint
  {http://arxiv.org/abs/1504.08280} {arXiv:1504.08280 [hep-lat]} \BibitemShut
  {NoStop}%
\bibitem [{\citenamefont {D'Elia}\ \emph {et~al.}(2022)\citenamefont {D'Elia},
  \citenamefont {Maio}, \citenamefont {Sanfilippo},\ and\ \citenamefont
  {Stanzione}}]{DElia:2021yvk}%
  \BibitemOpen
  \bibfield  {author} {\bibinfo {author} {\bibfnamefont {M.}~\bibnamefont
  {D'Elia}}, \bibinfo {author} {\bibfnamefont {L.}~\bibnamefont {Maio}},
  \bibinfo {author} {\bibfnamefont {F.}~\bibnamefont {Sanfilippo}}, \ and\
  \bibinfo {author} {\bibfnamefont {A.}~\bibnamefont {Stanzione}},\ }\href
  {\doibase 10.1103/PhysRevD.105.034511} {\bibfield  {journal} {\bibinfo
  {journal} {Phys. Rev. D}\ }\textbf {\bibinfo {volume} {105}},\ \bibinfo
  {pages} {034511} (\bibinfo {year} {2022})},\ \Eprint
  {http://arxiv.org/abs/2111.11237} {arXiv:2111.11237 [hep-lat]} \BibitemShut
  {NoStop}%
\bibitem [{\citenamefont {Hanel}\ and\ \citenamefont
  {Thurner}(2011{\natexlab{a}})}]{Hanel_2011_I}%
  \BibitemOpen
  \bibfield  {author} {\bibinfo {author} {\bibfnamefont {R.}~\bibnamefont
  {Hanel}}\ and\ \bibinfo {author} {\bibfnamefont {S.}~\bibnamefont
  {Thurner}},\ }\href {\doibase 10.1209/0295-5075/93/20006} {\bibfield
  {journal} {\bibinfo  {journal} {Europhysics Letters}\ }\textbf {\bibinfo
  {volume} {93}},\ \bibinfo {pages} {20006} (\bibinfo {year}
  {2011}{\natexlab{a}})}\BibitemShut {NoStop}%
\bibitem [{\citenamefont {Hanel}\ and\ \citenamefont
  {Thurner}(2011{\natexlab{b}})}]{Hanel_2011_II}%
  \BibitemOpen
  \bibfield  {author} {\bibinfo {author} {\bibfnamefont {R.}~\bibnamefont
  {Hanel}}\ and\ \bibinfo {author} {\bibfnamefont {S.}~\bibnamefont
  {Thurner}},\ }\href {\doibase 10.1209/0295-5075/96/50003} {\bibfield
  {journal} {\bibinfo  {journal} {Europhysics Letters}\ }\textbf {\bibinfo
  {volume} {96}},\ \bibinfo {pages} {50003} (\bibinfo {year}
  {2011}{\natexlab{b}})}\BibitemShut {NoStop}%
\bibitem [{\citenamefont {Nasser~Tawfik}(2016)}]{NasserTawfik:2016sqs}%
  \BibitemOpen
  \bibfield  {author} {\bibinfo {author} {\bibfnamefont {A.}~\bibnamefont
  {Nasser~Tawfik}},\ }\href {\doibase 10.1140/epja/i2016-16253-4} {\bibfield
  {journal} {\bibinfo  {journal} {Eur. Phys. J. A}\ }\textbf {\bibinfo {volume}
  {52}},\ \bibinfo {pages} {253} (\bibinfo {year} {2016})},\ \Eprint
  {http://arxiv.org/abs/1607.01264} {arXiv:1607.01264 [nucl-th]} \BibitemShut
  {NoStop}%
\bibitem [{\citenamefont {Tawfik}\ \emph {et~al.}(2017)\citenamefont {Tawfik},
  \citenamefont {Yassin},\ and\ \citenamefont {Abo~Elyazeed}}]{Tawfik:2017bul}%
  \BibitemOpen
  \bibfield  {author} {\bibinfo {author} {\bibfnamefont {A.~N.}\ \bibnamefont
  {Tawfik}}, \bibinfo {author} {\bibfnamefont {H.}~\bibnamefont {Yassin}}, \
  and\ \bibinfo {author} {\bibfnamefont {E.~R.}\ \bibnamefont {Abo~Elyazeed}},\
  }\href {\doibase 10.1088/1674-1137/41/5/053107} {\bibfield  {journal}
  {\bibinfo  {journal} {Chin. Phys. C}\ }\textbf {\bibinfo {volume} {41}},\
  \bibinfo {pages} {053107} (\bibinfo {year} {2017})},\ \Eprint
  {http://arxiv.org/abs/1701.04697} {arXiv:1701.04697 [nucl-th]} \BibitemShut
  {NoStop}%
\bibitem [{\citenamefont {Tawfik}\ \emph
  {et~al.}(2018{\natexlab{c}})\citenamefont {Tawfik}, \citenamefont {Yassin},\
  and\ \citenamefont {Abo~Elyazeed}}]{Tawfik:2018ahq}%
  \BibitemOpen
  \bibfield  {author} {\bibinfo {author} {\bibfnamefont {A.~N.}\ \bibnamefont
  {Tawfik}}, \bibinfo {author} {\bibfnamefont {H.}~\bibnamefont {Yassin}}, \
  and\ \bibinfo {author} {\bibfnamefont {E.~R.}\ \bibnamefont {Abo~Elyazeed}},\
  }\href {\doibase 10.1007/s12648-018-1216-2} {\bibfield  {journal} {\bibinfo
  {journal} {Indian J. Phys.}\ }\textbf {\bibinfo {volume} {92}},\ \bibinfo
  {pages} {1325} (\bibinfo {year} {2018}{\natexlab{c}})},\ \Eprint
  {http://arxiv.org/abs/1802.04912} {arXiv:1802.04912 [nucl-th]} \BibitemShut
  {NoStop}%
\bibitem [{\citenamefont {Tawfik}(2018)}]{Tawfik:2017bsy}%
  \BibitemOpen
  \bibfield  {author} {\bibinfo {author} {\bibfnamefont {A.~N.}\ \bibnamefont
  {Tawfik}},\ }\href {\doibase 10.1134/S1547477118030196} {\bibfield  {journal}
  {\bibinfo  {journal} {Phys. Part. Nucl. Lett.}\ }\textbf {\bibinfo {volume}
  {15}},\ \bibinfo {pages} {199} (\bibinfo {year} {2018})},\ \Eprint
  {http://arxiv.org/abs/1712.04807} {arXiv:1712.04807 [hep-ph]} \BibitemShut
  {NoStop}%
\bibitem [{\citenamefont {Rath}\ and\ \citenamefont
  {Dash}(2023)}]{Rath:2023bmx}%
  \BibitemOpen
  \bibfield  {author} {\bibinfo {author} {\bibfnamefont {S.}~\bibnamefont
  {Rath}}\ and\ \bibinfo {author} {\bibfnamefont {S.}~\bibnamefont {Dash}},\
  }\href {\doibase 10.1140/epjc/s10052-023-12051-3} {\bibfield  {journal}
  {\bibinfo  {journal} {Eur. Phys. J. C}\ }\textbf {\bibinfo {volume} {83}},\
  \bibinfo {pages} {867} (\bibinfo {year} {2023})},\ \Eprint
  {http://arxiv.org/abs/2303.03071} {arXiv:2303.03071 [hep-ph]} \BibitemShut
  {NoStop}%
\bibitem [{\citenamefont {Rath}\ and\ \citenamefont
  {Dash}(2024)}]{Rath:2023abm}%
  \BibitemOpen
  \bibfield  {author} {\bibinfo {author} {\bibfnamefont {S.}~\bibnamefont
  {Rath}}\ and\ \bibinfo {author} {\bibfnamefont {S.}~\bibnamefont {Dash}},\
  }\href {\doibase 10.1140/epja/s10050-024-01252-3} {\bibfield  {journal}
  {\bibinfo  {journal} {Eur. Phys. J. A}\ }\textbf {\bibinfo {volume} {60}},\
  \bibinfo {pages} {29} (\bibinfo {year} {2024})},\ \Eprint
  {http://arxiv.org/abs/2307.12002} {arXiv:2307.12002 [hep-ph]} \BibitemShut
  {NoStop}%
\bibitem [{\citenamefont {Islam}\ \emph {et~al.}(2021)\citenamefont {Islam},
  \citenamefont {Dey},\ and\ \citenamefont {Ghosh}}]{Islam:2019tlo}%
  \BibitemOpen
  \bibfield  {author} {\bibinfo {author} {\bibfnamefont {C.~A.}\ \bibnamefont
  {Islam}}, \bibinfo {author} {\bibfnamefont {J.}~\bibnamefont {Dey}}, \ and\
  \bibinfo {author} {\bibfnamefont {S.}~\bibnamefont {Ghosh}},\ }\href
  {\doibase 10.1103/PhysRevC.103.034904} {\bibfield  {journal} {\bibinfo
  {journal} {Phys. Rev. C}\ }\textbf {\bibinfo {volume} {103}},\ \bibinfo
  {pages} {034904} (\bibinfo {year} {2021})},\ \Eprint
  {http://arxiv.org/abs/1901.09543} {arXiv:1901.09543 [nucl-th]} \BibitemShut
  {NoStop}%
\bibitem [{\citenamefont {Li}\ \emph {et~al.}(2019)\citenamefont {Li},
  \citenamefont {Xu}, \citenamefont {Wang},\ and\ \citenamefont
  {Huang}}]{Li:2018ygx}%
  \BibitemOpen
  \bibfield  {author} {\bibinfo {author} {\bibfnamefont {Z.}~\bibnamefont
  {Li}}, \bibinfo {author} {\bibfnamefont {K.}~\bibnamefont {Xu}}, \bibinfo
  {author} {\bibfnamefont {X.}~\bibnamefont {Wang}}, \ and\ \bibinfo {author}
  {\bibfnamefont {M.}~\bibnamefont {Huang}},\ }\href {\doibase
  10.1140/epjc/s10052-019-6703-x} {\bibfield  {journal} {\bibinfo  {journal}
  {Eur. Phys. J. C}\ }\textbf {\bibinfo {volume} {79}},\ \bibinfo {pages} {245}
  (\bibinfo {year} {2019})},\ \Eprint {http://arxiv.org/abs/1801.09215}
  {arXiv:1801.09215 [hep-ph]} \BibitemShut {NoStop}%
\end{thebibliography}%

\end{document}